\documentclass[11pt,a4paper]{article}

\usepackage{amssymb}
\usepackage{amsmath}
\usepackage{amsfonts}
\usepackage{authblk}
\usepackage{setspace}
\usepackage{color}
\usepackage[pdftex]{graphicx} 

\usepackage{placeins}

\numberwithin{equation}{section}

\newcommand{\BbbR}{\mathbb{R}}
\newcommand{\BbbZ}{\mathbb{Z}}

\DeclareMathOperator{\sgn}{sgn}

\DeclareMathOperator{\arsinh}{arsinh}

\title{Unruh and analogue Unruh temperatures for\\ 
circular motion in 3+1 and 2+1 dimensions}

\author[1]{Steffen Biermann\footnote{steffen.biermann@nottingham.ac.uk}}
\author[1,2,3]{Sebastian Erne\footnote{erne@atomchip.org}}
\author[1]{Cisco Gooding\footnote{cisco.gooding@nottingham.ac.uk}}
\author[1]{Jorma Louko\footnote{jorma.louko@nottingham.ac.uk}}
\author[2]{J\"org Schmiedmayer\footnote{schmiedmayer@atomchip.org}}
\author[4,5]{William G. Unruh\footnote{unruh@physics.ubc.ca}}
\author[1,6]{Silke Weinfurtner\footnote{silke.weinfurtner@nottingham.ac.uk}}

\affil[1]{School of Mathematical Sciences, 
University of Nottingham,
Nottingham NG7 2RD, 
UK}

\affil[2]{Vienna Center for Quantum Science and Technology, Atominstitut, TU Wien, Stadionallee 2, A-1020, Vienna, Austria} 

\affil[3]{Wolfgang Pauli Institut, c/o Fak.\ Mathematik, Universit{\"a}t Wien,
Nordbergstrasse 15, 1090 Vienna, Austria}

\affil[4]{Department of Physics and Astronomy, The University of British Columbia, Vancouver, Canada V6T 1Z1} 

\affil[5]{Hagler Institute for Advanced Study and Institute for Quantum Science and Engineering, Texas A\&M University, College Station, TX, 77843-4242, USA}

\affil[6]{Centre for the Mathematics and Theoretical Physics of Quantum Non-Equilibrium Systems, 
University of Nottingham,
Nottingham NG7 2RD, 
UK}

\date{{\small July 2020; Revised September 2020\footnote{Post-publication update March 2023: typo in \eqref{eq:T/Tlin:largegap} corrected.}\\[1ex]
Published in Phys.\ Rev.\ D {\bf 102}, 085006 (2020)}}

\begin{document}

\maketitle
\begin{abstract}
The Unruh effect states that a uniformly linearly accelerated observer with proper acceleration $a$ experiences Minkowski vacuum as a thermal state in the temperature $T_{\text{lin}} = a/(2\pi)$, operationally measurable via the detailed balance condition between excitation and de-excitation probabilities. An observer in uniform circular motion experiences a similar Unruh-type temperature~$T_{\text{circ}}$, operationally measurable via the detailed balance condition, but $T_{\text{circ}}$ depends not just on the proper acceleration but also on the orbital radius and on the excitation energy. We establish analytic results for $T_{\text{circ}}$ for a massless scalar field in $3+1$ and $2+1$ spacetime dimensions in several asymptotic regions of the parameter space, and we give numerical results in the interpolating regions. In the ultrarelativistic limit, we verify that in $3+1$ dimensions $T_{\text{circ}}$ is of the order of $T_{\text{lin}}$ uniformly in the energy, as previously found by Unruh, but in $2+1$ dimensions $T_{\text{circ}}$ is significantly lower at low energies. We translate these results to an analogue spacetime nonrelativistic field theory in which the circular acceleration effects may become experimentally testable in the near future. We establish in particular that the circular motion analogue Unruh temperature grows arbitrarily large in the near-sonic limit, encouragingly for the experimental prospects, but the growth is weaker in effective spacetime dimension $2+1$ than in $3+1$. 
\end{abstract}


\section{Introduction}

The Unruh effect \cite{Fulling:1972md,Davies:1974th,Unruh:1976db} 
states that a linearly uniformly accelerated observer in Minkowski spacetime 
reacts to a quantum field in its Minkowski vacuum 
by excitations and de-excitations with the characteristics of a thermal state in 
the Unruh temperature~$a\hbar/(2\pi c k_B)$, 
where $a$ is the observer's proper acceleration 
(for textbooks and reviews, 
see~\cite{Birrell:1982ix,Wald:1995yp,Crispino:2007eb,Fulling:2014wzx}). 
A~direct experimental confirmation of the effect has however remained elusive 
because of the required magnitude of the acceleration~\cite{Fulling:2014wzx}. 
Prospects to observe versions of the effect in high-power 
laser systems are discussed in 
\cite{Chen:1998kp,Mourou:2006zz,Schutzhold:2006gj,Brodin:2007vf,gregori-scirep,Chen:2015bcg,Chen:2020sir}, 
and a selection of other experimental proposals are discussed in 
\cite{Alsing:2005dno,Retzker-circular,MartinMartinez:2010sg,Rodriguez-Laguna:2016kri,Kosior:2018vgx,Adjei:2020fmr}. 
Within the analogue spacetime programme of simulating relativistic phenomena 
in nonrelativistic laboratory systems~\cite{Barcelo:2005fc,Volovik-universe-2003}, 
an indirect experiment relying on virtual observers was reported 
in \cite{Leonhardt:2017lwm} and  
an indirect experiment relying on functional equivalence was reported in~\cite{Hu:2018psq}. 
A~direct experimental confirmation 
would have intrinsic interest as a demonstration of quantum vacuum friction,
as well as broader interest 
because of the connections to the Hawking effect~\cite{Hawking:1974sw}, 
and because of the connections 
to the early universe quantum effects 
that may be responsible for the origin 
of structure in the present-day Universe~\cite{Parker:1969au,mukha-wini-book}. 

While the Unruh effect in its standard form concerns 
uniform linear acceleration in Minkowski spacetime, 
similar phenomena exist also for other spacetimes and other motions. A~well-known example is 
the Gibbons-Hawking effect for inertial motion in de~Sitter spacetime~\cite{Gibbons:1977mu}, 
for which an analogue spacetime simulation has been proposed in~\cite{Fedichev:2003id,Fedichev:2003dj}. 
In Minkowski spacetime, a similar effect exists for uniform 
accelerations that are not linear~\cite{Letaw:1980yv,Korsbakken:2004bv,Good:2020hav}, 
including uniform circular 
motion~\cite{Letaw:1979wy,Takagi:1986kn,Doukas:2010wt,Jin:2014coa,Jin:2014spa}. 
The circular motion version is related to 
spin depolarisation in accelerator storage 
rings~\cite{Bell:1982qr,Bell:1986ir,Leinaas:1998tu,unruhcirclong,unruhcircshort}, which was 
originally predicted by different methods \cite{Sokolov:1963zn,Jackson:1975qi}, and which has 
been observed~\cite{Johnson:1982rb}, 
but the relation between this 
observation and the circular motion 
Unruh effect remains indirect~\cite{unruhcirclong,unruhcircshort}. 
A~proposal to observe the circular motion version 
in an electromagnetic cavity is discussed in~\cite{Lochan:2019osm}. 

Experimental interest in the circular motion Unruh effect in Minkowski spacetime has been recently reinvigorated by the experimental proposals 
put forward in \cite{Retzker-circular} and~\cite{unruh-rqin19talk,Gooding:2020scc}, 
within the analogue spacetime programme of simulating relativistic phenomena in nonrelativistic laboratory systems~\cite{Barcelo:2005fc,Volovik-universe-2003}. 
Among the four types of uniform nonlinear acceleration that exist in four-dimensional Minkowski spacetime \cite{Letaw:1980yv,Korsbakken:2004bv,Good:2020hav}, circular acceleration has two unique advantages over linear acceleration. 
First, circular motion allows the accelerating system to remain 
within a finite-size laboratory for an arbitrarily long interaction time. 
Second, in uniform circular motion in Minkowski spacetime, the time dilation 
gamma-factor between Minkowski time and the worldline's proper time remains constant 
over the worldline, unlike what happens in
uniform linear acceleration, or in any of the other nonlinear uniform accelerations \cite{Letaw:1980yv,Korsbakken:2004bv,Good:2020hav}. Modelling circular motion time dilation in a condensed matter system can therefore be accomplished simply by scaling the energies in the theoretical analysis of the experiment by a time-independent gamma-factor. By contrast, modelling time dilation in linear 
acceleration, or in any of the other nonlinear accelerations, would need to involve exponentially time-dependent energy scalings \cite{Retzker-circular,Olson:2010jy,Olson:2011bq}, raising the problem of engineering such time-dependent scalings in the condensed matter system, and limiting the time for which the system can be kept in the linear dispersion relation regime in which the analogue spacetime correspondence operates. 

Experimental proposals based on the circular motion Unruh effect have however also a disadvantage, 
both in genuinely relativistic systems and in analogue spacetime systems:  
the linear acceleration Unruh temperature formula no longer holds as an exact equality, 
and the actual Unruh temperature depends not just on the magnitude of the circular acceleration but also on the orbital speed and the energy at which the effect is probed. While the linear motion and circular motion temperatures are known to be of the same order of magnitude in certain regions of the parameter space~\cite{Good:2020hav,unruhcirclong,Muller:1995vk,Hodgkinson:2014iua}, 
a detailed control of their relation will be necessary for analysing prospective experiments. 

The purpose of this paper is to give a detailed comparison of the linear and circular motion Unruh temperatures, by a combination of analytic and numerical methods, in the case where the quantum field is a massless scalar field in Minkowski spacetime in its usual Minkowski vacuum state. We consider spacetime dimensions 3+1 and 2+1, as motivated by the experimental proposals. We address both a genuine relativistic spacetime system, which incorporates time dilation, and a condensed matter analogue spacetime system, in which the absence of time dilation is handled by a suitable energy scaling. 
We probe the field with a pointlike linearly-coupled Unruh-DeWitt detector 
\cite{Unruh:1976db,DeWitt:1979}, and we work in linear perturbation theory, 
in the limit of long interaction time but negligible back-action. 
The temperature seen by the detector will be defined operationally 
via the detailed balance condition between the excitation rate and the de-excitation rate. 
We leave it to future work to address effects due to other phenomena that will inevitably be present in experimental implementations, including 
finite size~\cite{Levin-Peleg-Peres,Davies:1996ks,Gutti:2010nv}, 
finite interaction time~\cite{Fewster:2016ewy}, 
nonzero ambient temperature~\cite{Hodgkinson:2014iua}, 
dispersion relation nonlinearity and Lorentz-noninvariance~\cite{Gutti:2010nv,Stargen:2017xii,Louko:2017emx}, 
and the detector's back-action on the field
\cite{Lin:2006jw,Moustos:2016lol,Sokolov:2018nmu}. 

For the $(3+1)$-dimensional relativistic system, 
we confirm that the circular motion Unruh temperature $T_{\text{circ}}$ 
agrees with the linear motion Unruh temperature within an energy-dependent factor of order unity in the ultrarelativistic limit, 
in agreement with the previous analytic scalar field results by Takagi~\cite{Takagi:1986kn}, 
M\"uller \cite{Muller:1995vk} 
and Unruh~\cite{unruhcirclong} 
(the published version \cite{unruhcircshort} of \cite{unruhcirclong} focused on the electromagnetic field), 
and consistently with the numerics given in 
\cite{Good:2020hav,unruhcirclong,Juarez-Aubry:2019gjw}. 
Beyond the ultrarelativistic limit the discrepancy is however larger, 
as we show by analytic results in several limits and by numerical results in the interpolating regions. 

For the $(2+1)$-dimensional relativistic system, $T_{\text{circ}}$ is 
qualitatively similar to that in 3+1 dimensions 
at high energies, but it is significantly smaller at low energies. 
In the ultrarelativistic limit, the 
$(2+1)$-dimensional $T_{\text{circ}}$ is suppressed at small energies 
relative to the $(3+1)$-dimensional value 
by the factor $1/\ln(1/|E|)$, where $E$ is the detector's energy gap. 

Results for the analogue spacetime system follow by scaling the relativistic energies 
by the time dilation gamma-factor. 
We find in particular that the temperature 
grows arbitrarily large in the near-sonic limit, 
encouragingly for the experimental prospects, but the 
growth is weaker in effective spacetime dimension 2+1 than in 3+1, 
by a factor proportional to $1/\ln\gamma$, where $\gamma$ is the time dilation gamma-factor.

We begin by recalling in Section \ref{sec:preliminaries}
relevant background about an Unruh-DeWitt detector in a relativistic spacetime, 
specialising to a stationary situation and reviewing the detailed balance definition of an effective temperature even when this temperature may depend on the energy, and finally specialising to uniform circular motion in Minkowski spacetime, with the quantum field in the Minkowski vacuum. 
Sections \ref{sec:3+1} and \ref{sec:2+1} address the relativistic system in respectively 3+1 and 2+1 dimensions. 
The translation to the analogue spacetime system is made in Section~\ref{sec:analogue}. Numerical plots are collected in Section~\ref{sec:numerical}. 
Section \ref{sec:conclusions} presents the conclusions and a discussion of the experimental upshots. 
Proofs of several technical results stated in the main text are deferred to five appendices. 

In the relativistic field theory we use units in which $c = \hbar = k_B = 1$, 
and in the analogue spacetime theory we use similar units in 
which the speed of sound has been set to unity. 
In asymptotic formulas, 
$O(x)$~denotes a quantity such that $O(x)/x$ is bounded as $x\to0$, 
$o(x)$ denotes a quantity such that $o(x)/x \to 0$ as $x\to0$, 
$O(1)$~denotes a quantity that remains bounded in the limit under consideration, 
and $o(1)$ denotes a quantity that goes to zero in the limit under consideration.

\section{Relativistic spacetime preliminaries\label{sec:preliminaries}}

In this section we review the relevant background about an Unruh-DeWitt detector coupled linearly to a scalar field in a relativistic spacetime. We first address a general stationary motion in a stationary quantum state, working in the limit of weak coupling and long interaction time but negligible back-action, 
and recalling how the detailed balance condition between the excitation rate and the de-excitation rate can be used to define an effective Unruh temperature even when this temperature depends on the energy of the transitions. We then specialise to a circular trajectory in Minkowski spacetime of dimension $d>2$ and to a massless scalar field prepared in its Minkowski vacuum state. 

\subsection{Field, detector, transition rate and temperature}

We consider a real Klein-Gordon scalar field $\phi$ 
in a relativistic spacetime, 
prepared initially in a quantum state denoted by~$|\Phi\rangle$. 
We assume the Wightman function 
$\mathcal{G}(\textsf{x},\textsf{x}') := \langle \Phi | \phi(\textsf{x})\phi(\textsf{x}') | \Phi\rangle$ 
to be a distribution with a sufficiently controlled singularity structure, 
including the Hadamard property at the coincidence limit $\textsf{x} \to \textsf{x}'$ \cite{Decanini:2005eg,Sanders:2013}. Further discussion about sufficient conditions is given in~\cite{Fewster:2016ewy}. 

We probe the field with an Unruh-DeWitt detector \cite{Unruh:1976db,DeWitt:1979}: 
a pointlike two-level quantum system on a prescribed smooth timelike trajectory $\textsf{x}(\tau)$, parametrised by the proper time~$\tau$. The detector's Hilbert space is spanned 
by the orthonormal basis $\lbrace |0 \rangle_D, |1 \rangle_D \rbrace$, 
such that 
$H_{\text{D}} |0\rangle_D = 0$ and $H_{\text{D}} |1 \rangle_D =
E |1 \rangle_D$, where $H_{\text{D}}$ is the detector's Hamiltonian 
with respect to $\tau$ and the constant $E\in\BbbR$ is the detector's energy gap. 
For $E > 0$ we may think of $|0 \rangle$ as the detector's ground state and
$|1 \rangle$ as the excited state; for $E < 0$ the roles are reversed. 

The interaction Hamiltonian is 
\begin{align}
H_{\text{int}}(\tau) = c \chi(\tau) \mu(\tau)
\phi\bigl(\textsf{x}(\tau)\bigr)
\ , 
\label{Hint}
\end{align}
where $c \in \BbbR$ is a coupling constant, 
$\mu$ is the detector's monopole moment
operator and $\chi$ 
is a real-valued smooth switching function that specifies how the 
interaction is turned on and off. In first-order perturbation theory, 
the probability for the detector to make a transition from $|0 \rangle_D$ 
to $|1 \rangle_D$, regardless the final state of~$\phi$, is 
\cite{Unruh:1976db,Birrell:1982ix,Wald:1995yp,DeWitt:1979}
\begin{align}
\mathcal{P}
= c^2 \left| \langle 1 | \mu(0) | 0 \rangle \right|^2 \mathcal{F}_{\chi}(E)
\ , 
\label{Probability}
\end{align}
where the (switching-dependent) response function $\mathcal{F}_\chi$ is given by 
\begin{align}
\mathcal{F}_{\chi}(E) := \int_{-\infty}^\infty \! d\tau' \, 
\int_{-\infty}^\infty \! d\tau'' \, \chi(\tau') \chi(\tau'') \, 
e^{-i E (\tau'-\tau'')} 
\, \mathcal{W}(\tau',\tau'')
\ , 
\label{ResponseFn}
\end{align}
$\mathcal{W(\tau,\tau')} := \mathcal{G}\bigl(\textsf{x}(\tau),\textsf{x}(\tau')\bigr)$ 
is the pull-back of the field's Wightman function to the detector's trajectory, 
and $\chi$ is assumed to have sufficiently strong early and late time falloff to make the integrals in \eqref{ResponseFn} convergent. 
Note that $\mathcal{W}$ is a distribution, with a coincidence limit singularity whose strength depends on the spacetime dimension, and it may have other singularities that depend on the details of the state~$|\Phi\rangle$, 
but under our assumptions about $\mathcal{G}$ these singularities are 
sufficiently controlled for the integrals in \eqref{ResponseFn} 
to exist~\cite{Hormander:1983,Fewster:1999gj}. 
Note also that $\mathcal{F}_\chi$ is manifestly real-valued because 
$\mathcal{W(\tau,\tau')} = \overline{\mathcal{W(\tau',\tau)}}$, 
where the overline denotes complex conjugation. 

The key point here is that the response function $\mathcal{F}_{\chi}$ 
\eqref{ResponseFn}
encodes how the detector's transition probability depends on the 
field's initial state and on the detector's energy gap, 
trajectory and switching. 
The detector's internal structure and the coupling strength 
enter only via the constant overall factor in~\eqref{Probability}. 
This constant overall factor will not play a role in what follows. 

We now specialise to the situation where both the trajectory and the state $|\Phi\rangle$ 
are stationary, in the sense that $\mathcal{W}$ depends on its two arguments only through their difference, 
\begin{align}
\mathcal{W}(\tau',\tau'') = \mathcal{W}(\tau'-\tau'',0)
\ . 
\label{eq:Wstat-def}
\end{align}
The only time dependence in the detector's response 
comes then from the switching function~$\chi$. 
We further specialise to the limit in which the detector operates for a long time, 
while the coupling nevertheless is so weak
that first-order perturbation theory remains applicable. 
Dividing $\mathcal{F}_{\chi}$ \eqref{ResponseFn} 
by the total interaction time and letting this interaction time tend to infinity
shows that
the transition probability per unit time, or the transition rate, 
is proportional to the (stationary) response function~${\mathcal F}$, 
given by \cite{Birrell:1982ix,Wald:1995yp}
\begin{align}
{\mathcal F}(E) := 
\int_{-\infty}^{\infty} ds \, e^{-iEs} \, {\mathcal W}(s,0)
\ . 
\label{eq:respfunc-general}
\end{align}
The sense of the long time limit in the passage to \eqref{eq:respfunc-general} has significant subtlety, 
including the validity of linear perturbation theory, 
the uniformity of the long time limit in $E$~\cite{Fewster:2016ewy}, 
and the relation to the occupation numbers 
in a stationary state experiment~\cite{Juarez-Aubry:2019gjw}, 
all of which would need to be addressed in concrete experimental proposals. 
In this paper we work within the idealised regime in which the transition rate is 
stationary and a multiple of ${\mathcal F}$~\eqref{eq:respfunc-general}. 

We define the Unruh temperature $T$ as seen by the detector by assuming that the detector's excitation and de-excitation rates are related by Einstein's detailed balance condition ${\mathcal F}(-E) = e^{E/T}{\mathcal F}(E)$ \cite{Einstein:detailed}, from which 
\begin{align}
\frac{1}{T} = \frac{1}{E} \ln\!\left(\frac{{\mathcal F}(-E)}{{\mathcal F}(E)}\right)
\ . 
\label{eq:detailedbalance-temperature} 
\end{align}
For a conventional thermal state, $T$ is independent of~$E$, 
as follows from the imaginary time periodicity of the Wightman function known as the Kubo-Martin-Schwinger
condition \cite{Kubo:1957mj,Martin:1959jp,Haag:1967sg}, 
and as is reviewed for an Unruh-DeWitt type detector in \cite{Takagi:1986kn,Fredenhagen:1986jg}. 
This is in particular the case for 
uniform linear acceleration in Minkowski vacuum, 
where $T$ is independent of $E$ and equal to $a/(2\pi)$, 
with $a$ being the magnitude of the proper acceleration~\cite{Unruh:1976db}: 
this is the usual Unruh effect. 
We consider situations where $T$ may depend on~$E$.

\subsection{Spacetime, field state and detector trajectory\label{subsec:trajectory}}

We specialise to Minkowski spacetime of dimension $d>2$, 
and to a massless scalar field that is initially prepared in its usual Minkowski vacuum. 
We use standard Minkowski coordinates in which 
$\textsf{x} = (t,\textbf{x}) = (t,x^1,\ldots,x^{d-1})$ 
and the metric reads 
\begin{align}
ds^2 &= 
- dt^2 + (d\textbf{x})^2 
\notag
\\
&= 
- dt^2 + {(dx^1)}^2 + \cdots + {(dx^{d-1})}^2
\ . 
\label{eq:Minkmetric}
\end{align}
The Wightman function is given by
\begin{align}
\label{eq:scalar-wightman-massless}
\mathcal{G}(\textsf{x},\textsf{x}')
= 
\frac{\Gamma\bigl(\frac{d}{2}-1\bigr)}{4\pi^{d/2} \bigl[{(\textbf{x}-\textbf{x}')}^2 - {(t-t'-i\epsilon)}^2\bigr]^{(d-2)/2}}
\ , 
\end{align}
where the distributional limit $\epsilon \to 0_+$ is understood, 
and the overall phase and the locus of $i\epsilon$ 
have been adjusted from the Feynman propagator analysis of \cite{Decanini:2005eg} 
to the Wightman two-point function~\cite{Kay:1988mu}. 
For odd~$d$, the denominator in \eqref{eq:scalar-wightman-massless} is positive for spacelike separations and the $i\epsilon$ specifies the branch on continuation to timelike separations. 

We take the detector to be in uniform circular motion. 
The worldline is 
\begin{align}
{\sf{x}}(\tau) =
\bigl(\gamma \tau, R \cos(\gamma\Omega\tau),  R \sin(\gamma\Omega\tau), \cdots \bigr)
\ , 
\label{eq:helical-trajectory}
\end{align}
where the dots are absent in $2+1$ spacetime dimensions and stand for the requisite number of zeroes in higher spacetime dimensions. 
$R$~and $\Omega$ are positive parameters satisfying $R\Omega < 1$, 
and $\gamma = 1/\sqrt{1 - R^2 \Omega^2}$.  
$R$~is the radius of the orbit, $\Omega$ is the angular velocity with respect to Minkowski time~$t$, 
and $\tau$ is the proper time. 
The orbital speed with respect to Minkowski time $t$ is $v = R\Omega$. 
The proper acceleration has magnitude $a = \sqrt{(\ddot{\sf{x}})^2} = R\Omega^2\gamma^2 = R^{-1}v^2\gamma^2$, where the overdot stands for derivative with respect to~$\tau$. 
Note that the orbital speed is constant over the detector's worldline: 
this is a crucial difference between circular acceleration and linear acceleration. 

We adopt $R$ and $v$ as a pair of independent parameters that specify the trajectory. 
It follows that 
\begin{subequations}
\begin{align}
\Omega &= v/R
\ , 
\\
\gamma &= 1/\sqrt{1 - v^2}
\ ,
\\
a &= 
R^{-1} v^2/(1-v^2)
\ ,  
\label{eq:a-formula}
\end{align}
\end{subequations}
and 
\begin{align}
\bigl(\Delta {\sf{x}}(\tau)\bigr)^2
&:= \bigl( {\sf{x}}(\tau) - {\sf{x}}(0) \bigr)^2
\notag 
\\
&= - 4 R^2 \left( \frac{z^2}{v^2} - \sin^2\! z \right)
\ , 
\label{eq:Deltax2}
\end{align}
where 
$z 
= 
\left(\frac{\gamma v}{2R}\right) \! \tau$. 

Using \eqref{eq:Wstat-def}, 
\eqref{eq:scalar-wightman-massless} 
and~\eqref{eq:Deltax2}, 
the stationary response function $\mathcal{F}$ is given by \eqref{eq:respfunc-general} where 
\begin{align}
{\mathcal W}(s,0)
= 
\frac{\Gamma\bigl(\frac{d}{2}-1\bigr)}{4\pi^{d/2} \bigl[\bigl(\Delta {\sf{x}}(s - i \epsilon)\bigr)^2\bigr]^{(d-2)/2}}
\ , 
\label{eq:W-circular-gendim}
\end{align}
understood in the sense of the distributional limit $\epsilon\to0_+$. 
For odd~$d$, the denominator in \eqref{eq:W-circular-gendim} has the phase of $i^{d-2}$ for $s>0$ and the phase of $(-i)^{d-2}$ for $s<0$.

\subsection{Circular temperature versus linear temperature}

Collecting, the Unruh temperature $T_{\text{circ}}$ for uniform circular motion is given by 
\eqref{eq:respfunc-general} and 
\eqref{eq:detailedbalance-temperature} 
with \eqref{eq:Deltax2} and~\eqref{eq:W-circular-gendim}. 

By comparison, recall that the Unruh temperature for uniform linear acceleration with proper 
acceleration $a$ is equal to $a/(2\pi)$~\cite{Unruh:1976db}. 
If the same were true for uniform circular motion, formula 
\eqref{eq:a-formula} would predict the Unruh temperature 
\begin{align}
T_{\text{lin}} &= \frac{a}{2\pi} = \frac{v^2}{2\pi (1-v^2) R}
\ . 
\label{eq:Tlin-gendef}
\end{align}
Our aim is first to investigate how the linear motion prediction \eqref{eq:Tlin-gendef} 
compares to the actual circular motion temperature 
in four and three spacetime dimensions, 
and then to translate these results into the corresponding analogue spacetime setting.

\section{3+1 dimensions\label{sec:3+1}} 

In this section we address analytically the relativistic theory in four spacetime dimensions, $d=4$. 
We first isolate the contributions to the response function from the 
distributional and non-distributional 
parts, by applying the arbitrary worldline result given in \cite{Louko:2006zv,Satz:2006kb} to circular motion. 
Recent applications of this isolation technique to circular motion appear in~\cite{Good:2020hav,Hodgkinson:2014iua}, 
an early application appears in Section 7.2 of~\cite{Takagi:1986kn}, 
and an application in the related context of vacuum fluctuations appears in~\cite{Letaw:1979wy}. 
By contrast, most of the previous work on circular motion uses for the 
response function and related quantities a distributional integral formula 
in which an $i\epsilon$ regulator is still present
\cite{Letaw:1980yv,Doukas:2010wt,Jin:2014coa,Jin:2014spa,Bell:1982qr,Bell:1986ir,Leinaas:1998tu,unruhcirclong,unruhcircshort,Lochan:2019osm,Muller:1995vk,Juarez-Aubry:2019gjw}, 
or eliminates the regulator by introducing a mode sum expansion~\cite{Levin-Peleg-Peres,Davies:1996ks,Gutti:2010nv,Stargen:2017xii}. 
We then give analytic results in three 
limits of interest. 
Numerical results will be given in Section~\ref{sec:numerical}. 

\subsection{Response function}

For $3+1$ dimensions, substituting $d=4$ in \eqref{eq:W-circular-gendim} gives 
\begin{align}
{\mathcal W}(s,0) = \frac{1}{4 \pi^2 \bigl(\Delta {\sf{x}}(s - i \epsilon)\bigr)^2}
\ , 
\end{align}
understood in the sense of the distributional limit $\epsilon\to0_+$. 
The only distributional contribution to the response function \eqref{eq:respfunc-general} 
comes from 
$s=0$. Isolating this contribution gives \cite{Louko:2006zv,Satz:2006kb} 
\begin{subequations}
\label{eq:F-combined}
\begin{align}
{\mathcal F}(E) &= {\mathcal F}^{\text{in}}(E) + {\mathcal F}^{\text{corr}}(E)
\ , 
\label{eq:F-split}
\\
{\mathcal F}^{\text{in}}(E) &= 
- \frac{E}{2\pi} \Theta(-E) 
\ , 
\label{eq:F-in}
\\
{\mathcal F}^{\text{corr}}(E) &= 
\frac{1}{2\pi^2} \int_0^\infty ds \cos(Es) 
\left(\frac{1}{s^2} + \frac{1}{\bigl(\Delta {\sf{x}}(s)\bigr)^2}\right)
\notag 
\\
&= 
\frac{1}{4\pi^2 \gamma v R} 
\int_0^\infty dz \, {\textstyle{\cos\!\left(\frac{2ER}{\gamma v} z\right)}} 
\left(\frac{\gamma^2 v^2}{z^2} - \frac{1}{z^2/v^2 - \sin^2 \! z} \right) 
\ , 
\label{eq:Fcorr-origint}
\end{align}
\end{subequations}
where $\Theta$ is the Heaviside function. 
${\mathcal F}^{\text{in}}$ is the inertial motion response function. 
Note that the integral in \eqref{eq:Fcorr-origint} 
has no singularities and converges in absolute value. 

While \eqref{eq:Fcorr-origint} is useful for numerical evaluation and for some analytic limits, 
an alternative that is useful for other analytic limits can be obtained as follows. 
Starting from~\eqref{eq:Fcorr-origint}, 
assuming $E\ne0$ and proceeding as in Appendix C of~\cite{Hodgkinson:2014iua}, we have 
\begin{align}
{\mathcal F}^{\text{corr}}(E) = 
- \frac{1}{8\pi^2 \gamma v R} 
\int_C dz \, \frac{\exp\!\left(i\frac{2|E|R}{\gamma v} z\right)}{z^2/v^2 - \sin^2 \! z} 
\ , 
\end{align}
where the contour $C$ is along the real axis from $-\infty$ to $\infty$ 
except for passing the pole at $z=0$ in the upper half-plane.
Closing the contour in the upper half-plane, the residue theorem gives 
\begin{subequations}
\begin{align}
{\mathcal F}^{\text{corr}}(E) 
&= {\mathcal F}^{\text{corr}}_0(E) + {\mathcal F}^{\text{corr}}_{+}(E)
\ , 
\\
{\mathcal F}^{\text{corr}}_0(E) &= 
\frac{\sqrt{\sinh^2\!\alpha_0 - \alpha_0^2} \, 
\exp\!\left(-\frac{2|E|R}{\gamma v} \alpha_0\right)}{8\pi R (\alpha_0 \cosh\alpha_0 - \sinh\alpha_0)\sinh\alpha_0}
\ , 
\label{eq:Fcorr0-def}
\\
{\mathcal F}^{\text{corr}}_{+}(E)
&= 
- \frac{v}{8\pi \gamma R}
\sum_{n\ne0} 
\frac{\exp\!\left(-\frac{2|E|R}{\gamma v} (\alpha_n + i \beta_n)\right)}{(\alpha_n +i \beta_n)
\left(1 - \alpha_n \coth\alpha_n - i \beta_n \tanh\alpha_n \right)} 
\ ,  
\label{eq:Fcorr+-def}
\end{align}
\end{subequations}
where $\alpha_n$ and $\beta_n$, $n\in\BbbZ$, 
are respectively the imaginary part and minus the real part of the zeroes of the function $z^2/v^2 - \sin^2 \! z$ in the upper half-plane, analysed in Appendix~\ref{app:auxiliary-zeroes}. 

We now turn to various limits of interest. 
The limits will be of the form where one variable is large or small while all other variables are held fixed, 
except for the ultrarelativistic limit $v\to1$ in subsection~\ref{subsec:3+1ultrarel}, 
which will be uniform in the remaining variables.

\subsection{Large gap limit}

Consider the limit $|E| \to \infty$ with fixed $v$ and~$R$.    
This is the ``low ambient temperature'' regime in 
Appendix C.3 of \cite{Hodgkinson:2014iua} and the results from there apply, 
as follows: 

By \eqref{eq:alphasequence}, ${\mathcal F}^{\text{corr}}(E) \to 0$, 
and its leading behaviour comes from ${\mathcal F}^{\text{corr}}_0(E)$. 
Using~\eqref{eq:detailedbalance-temperature}, 
the Unruh temperature is determined entirely by the 
coefficient of $|E|$ in the exponent in \eqref{eq:Fcorr0-def} 
and is given by 
\begin{align}
T_{\text{circ}} = \frac{\gamma v}{2\alpha_0 R} = \frac{1}{2 \sqrt{\sinh^2 \! \alpha_0 - \alpha_0^2} \, R}
\ . 
\label{eq:T-largegap}
\end{align}

By comparison, recall from \eqref{eq:Tlin-gendef} that 
the linear-motion-based prediction for the Unruh temperature is
\begin{align}
T_{\text{lin}} &= \frac{a}{2\pi} = \frac{v^2}{2\pi (1-v^2) R}
= \frac{\alpha_0^2}{2\pi (\sinh^2 \! \alpha_0 - \alpha_0^2) R}
\ . 
\label{eq:Tlin-circdef}
\end{align}
Hence 
\begin{align}
\frac{T_{\text{circ}}}{T_{\text{lin}}} = \frac{\pi \sqrt{\sinh^2 \! \alpha_0 - \alpha_0^2}}{\alpha_0^2}
\in \bigl(\pi/\sqrt{3} \,, \infty \bigr)
\ .
\label{eq:T/Tlin:largegap}
\end{align}
In the ultrarelativistic limit $v\to1$, 
we have $\alpha_0\to0$, and \eqref{eq:T/Tlin:largegap} gives 
\begin{align}
\frac{T_{\text{circ}}}{T_{\text{lin}}} = \frac{\pi}{\sqrt{3}}
\approx 
1.8
\ .   
\label{eq:T/Tlin:largegap-then-ultrarel}
\end{align}


\subsection{Small gap limit}

Consider the limit $E \to 0$ with fixed $v$ and~$R$.    

A dominated convergence argument in \eqref{eq:Fcorr-origint} 
shows that ${\mathcal F}^{\text{corr}}(E)$ is continuous in 
$E$ and the $E \to 0$ limit may be taken under the integral. 
Using \eqref{eq:detailedbalance-temperature}, 
\eqref{eq:F-combined} and \eqref{eq:Tlin-gendef}
gives 
\begin{align}
\frac{T_{\text{circ}}}{T_{\text{lin}}} = \frac{\sqrt{1-v^2}}{v} \int_0^\infty dz 
\left(\frac{1}{z^2} - \frac{1-v^2}{z^2- v^2 \sin^2 \! z} \right) 
\ . 
\label{eq:T/Tlin:zerogap}
\end{align}

\subsection{Ultrarelativistic limit\label{subsec:3+1ultrarel}}

Consider the ultrarelativistic limit $v\to1$. 
This is the limit in which the detector's 
orbital speed approaches the speed of light, 
in the Lorentz frame \eqref{eq:Minkmetric} introduced in subsection~\ref{subsec:trajectory}. 
We recall that the orbital speed is constant along the trajectory, unlike in linear acceleration. 

The ultrarelativistic limit was previously considered in Sections 7.2 and 7.3 of~\cite{Takagi:1986kn}, 
in~\cite{Muller:1995vk}, and in~\cite{unruhcirclong}. 
The core results for the temperature were found in formulas 
(12)--(14) of~\cite{unruhcirclong}. 
We reproduce these results here, in our equations 
\eqref{eq:T-ultrarel}--\eqref{eq:T/Tlin-ultrarel-largeandsmall} below, 
verifying in particular that the temperature result \eqref{eq:T-ultrarel} 
is uniform in the ratio~$E/a$. 

When $v\to1$, 
$\alpha_n$ and $\beta_n$ with $n\ne0$ 
tend to nonzero values as described in 
Appendix~\ref{app:auxiliary-zeroes}, 
and $\alpha_0$ tends to $0$ 
with the asymptotic behaviour~\eqref{eq:v-versus-alpha0:as}. 
Taking $\alpha_0$ as the independent parameter, we have 
\begin{subequations}
\begin{align}
v &= 1 - \tfrac16 \alpha_0^2 + O(\alpha_0^4)
\ ,
\\ 
a &= \frac{3}{\alpha_0^2 R} \bigl(1 + O(\alpha_0^2) \bigr)
\ ,
\label{eq:a-as-formula}
\\ 
\frac{1}{\gamma v} &= \frac{\alpha_0}{\sqrt{3}} \bigl(1 + O(\alpha_0^2) \bigr)
\ . 
\label{eq:gammav-inv-as-formula}
\end{align}
\end{subequations}
From \eqref{eq:Fcorr0-def}
and \eqref{eq:Fcorr+-def} we thus have 
\begin{subequations}
\label{eq:Fcorr-as}
\begin{align}
{\mathcal F}^{\text{corr}}_0(E) &= 
\frac{\sqrt{3} \exp\!\left(-\frac{2 \alpha_0 |E|R}{\gamma v}\right)}{8\pi R \alpha_0^2} 
\bigl(1 + O(\alpha_0^2) \bigr)
\ ,
\label{eq:Fcorr-as-0}
\\
{\mathcal F}^{\text{corr}}_{+}(E) &= 
O\biggl({\textstyle{\alpha_0\exp\!\left(-\frac{2 \alpha_1 |E|R}{\gamma v}\right)}}\biggr)
\ ,  
\label{eq:Fcorr-as-+}
\end{align}
\end{subequations}
in agreement 
with formula (11) in~\cite{Muller:1995vk}, 
using \eqref{eq:a-as-formula} and \eqref{eq:gammav-inv-as-formula}. 
The $O$-term in \eqref{eq:Fcorr-as-0} is independent of~$|E|R$, as seen from~\eqref{eq:Fcorr0-def}. 
The $O$-term in \eqref{eq:Fcorr-as-+} is uniform in~$|E|R$: 
this follows by a dominated convergence argument in~\eqref{eq:Fcorr+-def}, 
using the linear growth of $|\beta_n|$ established in Appendix~\ref{app:auxiliary-zeroes}. 

For the Unruh temperature, using \eqref{eq:detailedbalance-temperature}, 
keeping only the leading $v\to1$ behaviour, and expressing the result in terms of~$a$, we have
\begin{align}
T_{\text{circ}} = \frac{|E|}{\ln \biggl(1 + \frac{4\sqrt{3}\, |E|}{a} \exp\!\left(\frac{2\sqrt{3}\, |E|}{a}\right)\biggr)}
\ , 
\label{eq:T-ultrarel}
\end{align}
uniformly in $|E|/a$: this is equation (12) in~\cite{unruhcirclong}.  
The small and large $|E|/a$ limiting forms are 
\begin{subequations}
\label{eq:T-ultrarel-largeandsmall}
\begin{align}
T_{\text{circ}} &\approx \frac{a}{4\sqrt{3}} \ \ \ \ \text{for}\ \  |E|/a \ll 1
\ ,  
\label{eq:T-ultrarel-small}
\\[1ex]
T_{\text{circ}} &\approx \frac{a}{2\sqrt{3}}  \ \ \ \ \text{for}\ \  |E|/a \gg 1
\ ,  
\label{eq:T-ultrarel-large}
\end{align}
\end{subequations}
or 
\begin{subequations}
\label{eq:T/Tlin-ultrarel-largeandsmall}
\begin{align}
\frac{T_{\text{circ}}}{T_\text{lin}} &\approx \frac{\pi}{2\sqrt{3}} \approx 0.9 \ \ \ \ \text{for}\ \  |E|/a \ll 1
\ ,  
\label{eq:T/Tlin-ultrarel-small}
\\[1ex]
\frac{T_{\text{circ}}}{T_\text{lin}} &\approx \frac{\pi}{\sqrt{3}} \approx 1.8 \ \ \ \ \text{for}\ \  |E|/a \gg 1
\ .  
\label{eq:T/Tlin-ultrarel-large}
\end{align}
\end{subequations}
Note that \eqref{eq:T/Tlin-ultrarel-large}
agrees with~\eqref{eq:T/Tlin:largegap-then-ultrarel}.

\section{2+1 dimensions\label{sec:2+1}} 

In this section we address analytically the relativistic theory in three spacetime dimensions, $d=3$. 
Proceeding as in Section~\ref{sec:3+1}, 
we first isolate the contributions 
to the response function from the distributional and non-distributional parts 
of~${\mathcal W}$, by applying the arbitrary worldline result given in \cite{Hodgkinson:2011pc} to circular motion. 
To our knowledge this application has not been considered previously, although an alternative  
analytic expression for the response function as a mode sum has been given in~\cite{Louko:2017emx}. 
We then consider analytically four 
limits of interest. 
Numerical results will be given in Section~\ref{sec:numerical}.

\subsection{Response function}

For $2+1$ dimensions, substituting $d=3$ in \eqref{eq:W-circular-gendim} gives 
\begin{align}
{\mathcal W}(s,0) = 
\frac{1}{4 \pi} \times 
\frac{1}{\sqrt{\bigl(\Delta {\sf{x}}(s - i \epsilon)\bigr)^2}}
\ , 
\end{align}
understood in the sense of the distributional limit $\epsilon\to0_+$. 
The square root in the denominator is positive imaginary for $s>0$ and negative imaginary for $s<0$. 

The response function ${\mathcal F}$ \eqref{eq:respfunc-general} is discontinuous at zero argument. 
We assume throughout $E\ne0$. 

The only distributional contribution to the response function comes again from 
$s=0$. Isolating this contribution gives \cite{Hodgkinson:2011pc} 
\begin{align}
{\mathcal F}(E) = \frac{1}{4} - \frac{1}{2\pi}\int_0^\infty ds \, \frac{\sin(Es)}{\sqrt{-\bigl(\Delta {\sf{x}}(s)\bigr)^2}}
\ , 
\label{eq:Fint-2+1}
\end{align}
where the square root in the denominator is now positive. 
Using \eqref{eq:Deltax2}, this gives the split of ${\mathcal F}$ into its even and odd parts as 
\begin{subequations}
\label{eq:F2+1-first-split}
\begin{align}
{\mathcal F}(E) &= \frac{1}{4} + {\mathcal F}^{\text{odd}}(E)
\ , 
\label{eq:F2+1-eo-split}
\\
{\mathcal F}^{\text{odd}}(E)
&= 
- \frac{1}{2\pi \gamma v} 
\int_0^\infty dz \, \frac{\sin\!\left(\frac{2ER}{\gamma v} z\right)}{\sqrt{z^2/v^2 - \sin^2 \! z}} 
\ . 
\label{eq:F2+1-odd-def}
\end{align}
\end{subequations}
An alternative split is 
\begin{subequations}
\label{eq:F2+1-split-coll}
\begin{align}
{\mathcal F}(E) &= {\mathcal F}^{\text{in}}(E) + {\mathcal F}^{\text{corr}}(E)
\ , 
\label{eq:F2+1-split}
\\
{\mathcal F}^{\text{in}}(E) &= 
\tfrac12\Theta(-E) 
\ , 
\label{eq:F2+1-in}
\\
{\mathcal F}^{\text{corr}}(E) &= 
\frac{1}{2\pi} \int_0^\infty ds \sin(Es) 
\left(\frac{1}{s} - \frac{1}{\sqrt{-\bigl(\Delta {\sf{x}}(s)\bigr)^2}}\right)
\notag
\\
&= 
\frac{1}{2\pi \gamma v} 
\int_0^\infty dz \, {\textstyle{\sin\!\left(\frac{2ER}{\gamma v} z\right)}} 
\left(\frac{\gamma v}{z} - \frac{1}{\sqrt{z^2/v^2 - \sin^2 \! z}} \right) 
\ ,  
\label{eq:F2+1-corr}
\end{align}
\end{subequations}
where ${\mathcal F}^{\text{in}}$ is the inertial motion response function. 
All the integrals in \eqref{eq:Fint-2+1}, \eqref{eq:F2+1-odd-def} and \eqref{eq:F2+1-corr}
are free of singularities and 
converge as improper Riemann integrals. 

We note that ${\mathcal F}^{\text{corr}}$ is odd and 
\begin{align}
{\mathcal F}^{\text{corr}}(E) = \tfrac14 \sgn(E) + {\mathcal F}^{\text{odd}}(E)
\ . 
\label{eq:2+1:Fcorr-vs-Fodd}
\end{align}
Since ${\mathcal F}\ge0$ by construction, and since ${\mathcal F}^{\text{odd}}$ is odd, 
we have 
$|{\mathcal F}^{\text{odd}}| \le \frac14$, $0 \le {\mathcal F} \le \frac12$, 
and $|{\mathcal F}^{\text{corr}}| \le \frac12$. 
${\mathcal F}^{\text{corr}}(E)$ has the same sign as~$E$. 

We shall mainly work with \eqref{eq:F2+1-first-split} and~\eqref{eq:F2+1-split-coll}. 
We however record here two alternative expressions. 

First, starting from \eqref{eq:F2+1-corr} and proceeding as in Appendix C of \cite{Hodgkinson:2014iua} gives 
\begin{align}
{\mathcal F}^{\text{corr}}(E) = 
\frac{i \sgn(E)}{4\pi \gamma v} 
\int_C dz \, \frac{\exp\!\left(i\frac{2|E|R}{\gamma v} z\right)}{\sqrt{z^2/v^2 - \sin^2 \! z}} 
\ , 
\label{eq:F2+1corr:C-formula}
\end{align}
where the contour $C$ is along the real axis from $-\infty$ to $\infty$ 
except for passing the branch point at $z=0$ in the upper half-plane, and the square root 
in the denominator is positive for $z>0$ and negative for $z<0$. 
Deforming the contour to the upper half-plane gives 
\begin{subequations}
\label{eq:F0corr:2+1-combined}
\begin{align}
{\mathcal F}^{\text{corr}}(E) &= {\mathcal F}^{\text{corr}}_0(E) + {\mathcal F}^{\text{corr}}_{+}(E)
\ , 
\\
{\mathcal F}^{\text{corr}}_0(E) &= 
\frac{\sgn(E)}{2\pi \gamma v} \int_{\alpha_0}^\infty 
d\alpha \, 
\frac{\exp\!\left(-\frac{2|E|R}{\gamma v} \alpha\right)}{\sqrt{\sinh^2 \! \alpha - \alpha^2/v^2}} 
\ , 
\label{eq:F0corr:2+1}
\\
{\mathcal F}^{\text{corr}}_{+}(E)
&= 
- \frac{\sgn(E)}{2\pi\gamma} 
\sum_{n\ne0}  
\frac{i}{\beta_n} \, 
{\textstyle{\exp\!\left(-\frac{2|E|R}{\gamma v} (\alpha_n + i \beta_n)\right)}}
\int_0^\infty dy \, 
{\textstyle{\exp\!\left(-\frac{2|E|R}{\gamma v} y\right)}}
\notag
\\
&\hspace{3ex}
\times 
\Biggl\{ 
\frac{\cosh(\alpha_n+y)}{\cosh\alpha_n} + 1 
- \frac{i}{\beta_n} \left[\frac{\alpha_n}{\sinh\alpha_n} \sinh(\alpha_n+y) 
+ (\alpha_n+y) \right]
\Biggr\}^{-1/2}
\notag
\\
&\hspace{3ex}
\times 
\Biggl\{ 
\frac{\cosh(\alpha_n+y)}{\cosh\alpha_n} - 1 
- \frac{i}{\beta_n} \left[\frac{\alpha_n}{\sinh\alpha_n} \sinh(\alpha_n+y) 
- (\alpha_n+y) \right]
\Biggr\}^{-1/2}
\ , 
\label{eq:F+corr:2+1}
\end{align}
\end{subequations}
where $\alpha_n$ and $\beta_n$ are as given in Appendix~\ref{app:auxiliary-zeroes}. 
The square roots of a complex number in \eqref{eq:F+corr:2+1}
denote the branch with a positive real part. 
The convergence of the sum in \eqref{eq:F+corr:2+1} 
is however weaker than that of the corresponding four-dimensional sum~\eqref{eq:Fcorr+-def}, 
and this limits the usefulness of \eqref{eq:F0corr:2+1-combined} for analytic limits. 

Second, a mode sum expansion of the Wightman function yields for the response function the 
mode sum expression \cite{Louko:2017emx}
\begin{align}
{\mathcal F}(E) 
&= 
\frac{1}{2\gamma}
\sum_{m = \lceil ER/(v\gamma)\rceil}^\infty 
J_m^2\bigl(mv - (ER/\gamma)\bigr) 
\ , 
\label{eq:Fhelical2+1-normalfield}
\end{align}
where $\lceil \cdot \rceil$ is the ceiling function and $J_m$ 
is the Bessel function of the first kind. 
Formula \eqref{eq:Fhelical2+1-normalfield} is tractable numerically~\cite{Louko:2017emx}, 
but extracting analytic limits from it does not appear straightforward. 

We now turn to various limits of interest. 
All the limits will be of the form where one variable is large or small while all other variables are held fixed.

\subsection{Large gap limit}

Consider the limit $|E| \to \infty$ with fixed $v$ and~$R$.  

Proceeding as in $3+1$ dimensions, we find that the dominant contribution to 
${\mathcal F}^{\text{corr}}(E)$ comes from 
${\mathcal F}^{\text{corr}}_0(E)$. 
Changing the integration variable in \eqref{eq:F0corr:2+1} 
by $\alpha = \alpha_0(1+y^2)$ and using the stationary 
point expansion near $y=0$ \cite{wong} shows that this contribution is a multiple of 
$\sgn(E) |E|^{-1/2} \exp\!\left(-\frac{2|E|R}{\gamma v} \alpha_0\right)$, 
whose exponential factor is the same as in~\eqref{eq:Fcorr0-def}. 

The Unruh temperature is hence given by~\eqref{eq:T-largegap}, 
as in $3+1$ dimensions.

\subsection{Small gap limit\label{sec:2+1-smallgap}}

Consider the limit $E \to 0$ with fixed $v$ and~$R$.  

We show in Appendix \ref{app:2+1-smallgap} that 
\begin{align}
{\mathcal F}^{\text{corr}}(E) = \frac{\gamma-1}{4\gamma} \sgn(E) + O(E)
\ . 
\label{eq:Fcorr2+1-smallgap}
\end{align}
Using \eqref{eq:detailedbalance-temperature}, 
\eqref{eq:F2+1-split-coll} and~\eqref{eq:Fcorr2+1-smallgap}, we then have 
\begin{align}
T_{\text{circ}} = \frac{|E|}{\ln \bigl(\frac{\gamma+1}{\gamma-1}\bigr)}
\bigl(1 + O(E)\bigr)
\ , 
\label{eq:2+1-T-smallgap}
\end{align}
so that $T_{\text{circ}}\to0$ as $E\to0$. 
The temperature in this limit is hence significantly lower than the nonzero 
limit \eqref{eq:T/Tlin:zerogap} obtained in $3+1$ dimensions.

\subsection{Ultrarelativistic limit with fixed $E$\label{sec:2+1-ultrarel-fixedE}}

Analysing the ultrarelativistic limit $v\to1$ uniformly in $E/a$ is difficult 
because the convergence of \eqref{eq:F+corr:2+1} 
in this limit is weaker than the convergence of~\eqref{eq:Fcorr+-def}. 
In this subsection we consider the $v\to1$ limit with fixed $E$ and~$R$.   
The case of fixed $E/a$ 
will be addressed in subsection~\ref{sec:2+1-ultrarel}. 

We show in Appendix \ref{app:2+1-ultrarel-fixedE} that 
\begin{align}
{\mathcal F}^{\text{odd}}(E)
= - \frac{1}{4\gamma} \sgn(E) + o(1/\gamma)
\ . 
\label{eq:Fodd-2+1-fixedE-as}
\end{align}
By~\eqref{eq:F2+1-first-split}, this implies that ${\mathcal F}(E) \to \frac14$ as $v\to1$. 
(We note in passing that this is consistent with the numerical evidence shown  
in Figure 5 of~\cite{Louko:2017emx}, obtained by numerical evaluation from~\eqref{eq:Fhelical2+1-normalfield}.) 
From \eqref{eq:detailedbalance-temperature} 
and 
\eqref{eq:F2+1-eo-split} 
we then have 
\begin{align}
T_{\text{circ}} = \frac{\gamma |E|}{2}
\bigl(1 + o(1/\gamma)\bigr)
\ . 
\label{eq:2+1-T-ultrarel-fixedE}
\end{align}
Being proportional to $\gamma$, 
this temperature is significantly lower than the $3+1$ 
temperature shown in~\eqref{eq:T-ultrarel-small}, 
which is proportional to $\gamma^2$.


\subsection{Ultrarelativistic limit with fixed $E/a$\label{sec:2+1-ultrarel}}

Consider now the limit $v\to1$ with fixed~$E/a$. 

We verify in Appendix \ref{app:2+1-ultrarel-E/a-fixed} three properties. 
First, that 
\begin{subequations}
\begin{align}
&{\mathcal F}^{\text{odd}}(E) \to {\mathcal F}^{\text{odd}}_{\infty}(E)
= 
- \frac{1}{2\pi} G\bigl(2\sqrt{3}E/a\bigr)
\ , 
\label{eq:Fodd-2+1-limit}
\\
&{\mathcal F}^{\text{corr}}(E)
\to 
{\mathcal F}^{\text{corr}}_{\infty}(E)
= 
\frac{1}{2\pi} H\bigl(2\sqrt{3}E/a\bigr)
\ , 
\label{eq:Fcorr-2+1-limit}
\end{align}
\end{subequations}
where 
\begin{subequations}
\label{eq:GFfunc-def}
\begin{align}
G(q) &:= \int_0^\infty dx \, \frac{\sin(qx)}{x \sqrt{1+x^2}}
\ , 
\label{eq:Gfunc-def}
\\
H(q) &:= \sgn(q) \int_1^\infty dy \, \frac{e^{-|q|y}}{y\sqrt{y^2-1}}
\ .   
\label{eq:Hfunc-def}
\end{align}
\end{subequations}
Second, that 
the small argument asymptotic form of $G$ and the large argument asymptotic form of $H$ are respectively given by 
\begin{subequations}
\label{eq:GH-asymptotics}
\begin{align}
G(q)
&= 
q \ln(2e^{1-\gamma_E}/|q|) 
+ o(q)
\ , 
\label{eq:G-asymptotics}
\\
H(q) &= \sgn(q) 
\sqrt{\frac{\pi}{2|q|}} \, e^{-|q|} \Bigl(1 + O\bigl(|q|^{-1}\bigr) \Bigr)
\ ,  
\label{eq:H-asymptotics}
\end{align}
\end{subequations}
where $\gamma_E$ is Euler's constant. 
Third, that 
the functions $G$ and $H$ satisfy 
\begin{align}
H(q) = \frac{\pi}{2} \sgn(q) - G(q)
\ , 
\label{eq:H-G-relation}
\end{align}
so that 
\begin{align}
{\mathcal F}^{\text{corr}}_\infty(E) = \tfrac14 \sgn(E) + {\mathcal F}^{\text{odd}}_\infty(E)
\ , 
\label{eq:2+1:infty-Fcorr-vs-Fodd}
\end{align}
as must be for consistency with~\eqref{eq:2+1:Fcorr-vs-Fodd}. 

We note that 
Maple 2018 \cite{maple} gives for $H$ and $G$ expressions in terms of Meijer's $G$-function~\cite{dlmf}. 
We have used Maple numerical routines for these functions to make consistency 
checks of some of the numerical results of Section \ref{sec:numerical} below. 


The inverse Unruh temperature is given by 
\begin{align}
\frac{1}{T_{\text{circ}}} &= \frac{1}{|E|} \ln\!\left(\frac{1 + \frac{2}{\pi} G\bigl(2\sqrt{3}|E|/a\bigr)}{1 - \frac{2}{\pi} G\bigl(2\sqrt{3}|E|/a\bigr)}\right)
\notag
\\
&= 
\frac{1}{|E|} \ln\!\left(\frac{\pi - H\bigl(2\sqrt{3}|E|/a\bigr)}{H\bigl(2\sqrt{3}|E|/a\bigr)}\right)
\ . 
\label{eq:T-2+1-ultrarel}
\end{align}
By \eqref{eq:GH-asymptotics},
the small and large $|E|/a$ limiting forms are 
\begin{subequations}
\label{eq:T-2+1:ultrarel-largeandsmall}
\begin{align}
T_{\text{circ}} &\approx 
\frac{\pi a}{8\sqrt{3} \, \ln \Bigl( \frac{e^{1-\gamma_E}}{\sqrt{3}} \frac{a}{|E|} \Bigr)}
\ \ \ \ \text{for} \ \  |E|/a \ll 1
\ ,  
\label{eq:T-2+1:ultrarel-small}
\\
T_{\text{circ}} &\approx 
\frac{a}{2\sqrt{3}} 
\ \ \ \ \text{for} \ \ |E|/a \gg 1
\ .  
\label{eq:T-2+1:ultrarel-large}
\end{align}
\end{subequations}
Compared with the $3+1$ results \eqref{eq:T-ultrarel-largeandsmall}, 
the large $|E|/a$ regimes agree, 
but in the small $|E|/a$ regime the 2+1 temperature is suppressed 
by the logarithmic factor~$1/\ln(a/|E|)$.

\section{Analogue spacetime implementation\label{sec:analogue}} 

In this section we consider analogue spacetime 
implementations of the type proposed in~\cite{Retzker-circular} 
and~\cite{unruh-rqin19talk,Gooding:2020scc}, 
in a nonrelativistic condensed matter laboratory system, 
such as a Bose-Einstein condensate or superfluid helium~\cite{Barcelo:2005fc,Volovik-universe-2003}. 
We consider both effective spacetime dimension $2+1$ and effective spacetime dimension $3+1$. 

The condensed matter system provides an effective Minkowski 
geometry, in which the speed of light is replaced
by the speed of phonon-type excitations. 
We work in units in which this speed of sound is set to unity. 
The main new feature is that since the system has no analogue of relativistic time dilation, 
the energy of the moving detector is now defined with respect to the 
laboratory time, which is the Minkowski time in the effective 
Minkowski metric, and 
there is no analogue of a relativistic proper time. 
To maintain the analogue with the relativistic system, we assume that 
the detector's speed remains below the sonic limit $v=1$: 
we shall not consider nonlinear dispersion or 
analogue Cerenkov radiation~\cite{Marino:2020uqj}. 

We continue to consider an Unruh-DeWitt detector that is 
coupled linearly to the phonon-type quantum field. 
This is precisely the detector introduced in the proposal of \cite{Retzker-circular} 
to accelerate a quantum dot in a Bose-Einstein condensate. 
We note, however, that the results for the 
Unruh temperature will be independent of 
the detailed form of the coupling as long as the coupling is linear, 
given that we are working in the regime of long interaction but negligible back-action. 
This is because the detailed balance Unruh temperature 
depends on the detector's excitation and de-excitation rates only through their ratio. 
For example, if the coupling were not to the value of the field but to the time derivative of the field, 
as in the detection proposal of~\cite{unruh-rqin19talk,Gooding:2020scc}, 
or to higher time derivatives of the field, each time derivative would bring 
to the response function ${\mathcal F}(E)$ an additional factor~$E^2$, 
and these factors would just cancel from the temperature. 
More generally, any change in the coupling that affects 
both the excitation and de-excitation cross-sections in the same way, 
even if energy-dependent, will 
cancel out of the detailed balance Unruh temperature. 

We also assume the condensed matter system to be so large, 
compared with the parameters of the detector's orbit, 
that finite size effects remain negligible. 
It follows, as in the case of the relativistic field, 
that the circular motion 
Unruh effect seen by the detector does not 
have a description in terms of 
phonons adapted to a rigidly rotating quantisation frame: 
as the rotating frame has supersonic velocities 
sufficiently far from the centre of rotation, 
the frame does not provide a positive and negative frequency split 
on which a Fock quantisation of the field could be based, 
there is no notion of a `rotating vacuum' or `rotating phonons', 
and Bogoliubov coefficients cannot be introduced~\cite{Levin-Peleg-Peres,Davies:1996ks,Gutti:2010nv}. 

Under these assumptions, it is straightforward to translate 
our relativistic formalism to the laboratory setting, by writing 
\begin{subequations}
\label{eq:nonrel-acc-E-T}
\begin{align}
\hat E := E/\gamma
\ , 
\\
\hat T := T/\gamma
\ , 
\\
\hat a := a/\gamma^2
\ ,
\end{align}
\end{subequations}
where 
$\hat E$ is the energy gap with respect to the laboratory time~$t$,  
$\hat T$ is the temperature with respect to~$\hat E$, 
and 
$\hat a$ is the nonrelativistic acceleration. 
The linear-motion-based prediction for the analogue Unruh temperature is hence 
\begin{align}
{\hat T}_{\text{lin}} = T_{\text{lin}}/\gamma = \frac{a}{2\pi \gamma} 
= \frac{{\hat a} \gamma}{2\pi}
= \frac{\gamma v^2}{2\pi R}
\ ,  
\end{align}
while combining \eqref{eq:detailedbalance-temperature} and \eqref{eq:nonrel-acc-E-T}
shows that the actual analogue Unruh temperature is given by 
\begin{align}
\frac{1}{{\hat T}_{\text{circ}}} = 
\frac{1}{\hat E} 
\ln\!\left(\frac{{\mathcal F}(- \gamma \hat E)}{{\mathcal F}(\gamma \hat E)}\right)
\ , 
\label{eq:analogue-detailedbalance-temperature} 
\end{align}
where ${\mathcal F}$ is the relativistic response function found in
Sections \ref{sec:3+1} and~\ref{sec:2+1}. 

Numerical results are shown below in Section~\ref{sec:numerical}. We consider here 
analytically only the near-sonic limit 
$v\to1$ with $\hat a$ and $\hat E$ fixed. 
Then $E/a = ({\hat E}/{\hat a})/\gamma \to 0$ as $v\to1$. 
We show in Appendix \ref{app:analog-spacetime-asymptotics}
that 
\begin{subequations}
\label{eq:Tbecboth-ultrarel-small}
\begin{align}
{\hat T}_{\text{circ}} &\approx \frac{1}{4\sqrt{3}} \, \gamma \, {\hat a}
\ \ \ \ \text{in} \ \ 3+1
\ ,
\label{eq:Tbec-ultrarel-small}
\\[1ex]
{\hat T}_{\text{circ}} &\approx \frac{\pi}{8\sqrt{3}} 
\, \frac{\gamma}{\ln\gamma}
\, 
{\hat a}
\ \ \ \ \text{in} \ \ 2+1
\ ,  
\label{eq:Tbec-2+1:ultrarel-small}
\end{align}
\end{subequations}
or 
\begin{subequations}
\label{eq:Tratbecboth-ultrarel-small}
\begin{align}
\frac{{\hat T}_{\text{circ}}}{{\hat T}_{\text{lin}}} &\approx \frac{\pi}{2\sqrt{3}} 
\ \ \ \ \text{in} \ \ 3+1
\ ,
\label{eq:Ttratbec-ultrarel-small}
\\[1ex]
\frac{{\hat T}_{\text{circ}}}{{\hat T}_{\text{lin}}} &\approx \frac{\pi^2}{4\sqrt{3}} 
\, \frac{1}{\ln\gamma}
\ \ \ \ \text{in} \ \ 2+1
\ . 
\label{eq:Tratbec-2+1:ultrarel-small}
\end{align}
\end{subequations}
The temperature hence 
grows arbitrarily large 
in the $v\to1$ limit in both $3+1$ and $2+1$ dimensions, 
in $3+1$ dimensions proportionally to~$\gamma$, 
but in $2+1$ dimensions only proportionally to $\gamma/\ln\gamma$.

\begin{figure}[p]
\centering
\includegraphics[width=0.48\textwidth]{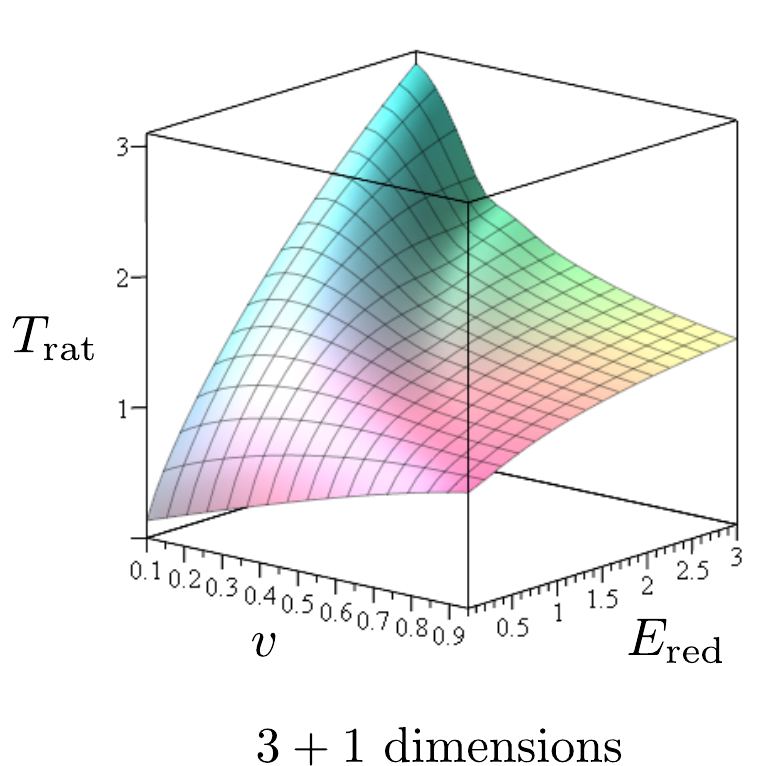} \hspace{2ex} \includegraphics[width=0.48\textwidth]{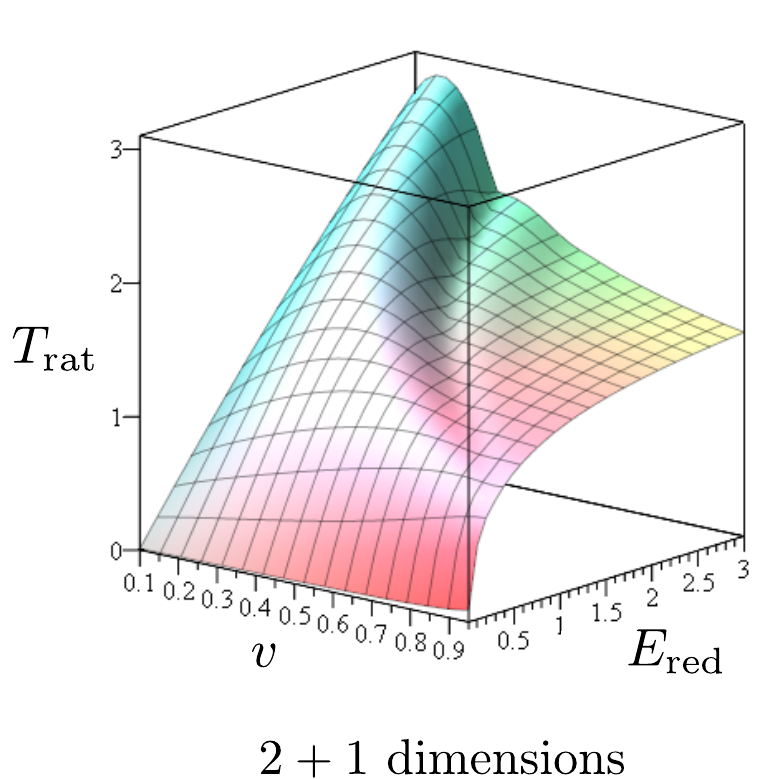}
\caption{Relativistic spacetime 
$T_{\text{rat}} := T_{\text{circ}}/T_{\text{lin}}$ 
as a function of $v$ and $E_{\text{red}} := E/a$, 
for $0.1 \le v \le 0.95$ and $0.1 \le E_{\text{red}} \le 3$. 
The plotting range was chosen for numerical stability, 
avoiding small and large values of $v$ and small and large values of $E_{\text{red}}$. 
Left in $3+1$ dimensions, evaluated from \eqref{eq:detailedbalance-temperature} 
with~\eqref{eq:F-combined}; 
right in $2+1$ dimensions, evaluated from \eqref{eq:detailedbalance-temperature} 
with~\eqref{eq:F2+1-split-coll}. 
In the limit $E_{\text{red}}\to0$, outside the plotted range, 
the $3+1$ graph tends to a nonzero value, as seen from~\eqref{eq:T/Tlin:zerogap}, 
while the $2+1$ graph has a significant drop, tending to zero linearly in~$E_{\text{red}}$, 
as seen from~\eqref{eq:2+1-T-smallgap}.
The continuations of the graphs to the ultrarelativistic 
limit $v\to1$, outside the plotted range, are shown in Figure~\ref{fig:ultrarel}.\label{fig:perspectiveTrat}} 
\end{figure}

\begin{figure}[p]
\centering
\includegraphics[width=0.48\textwidth]{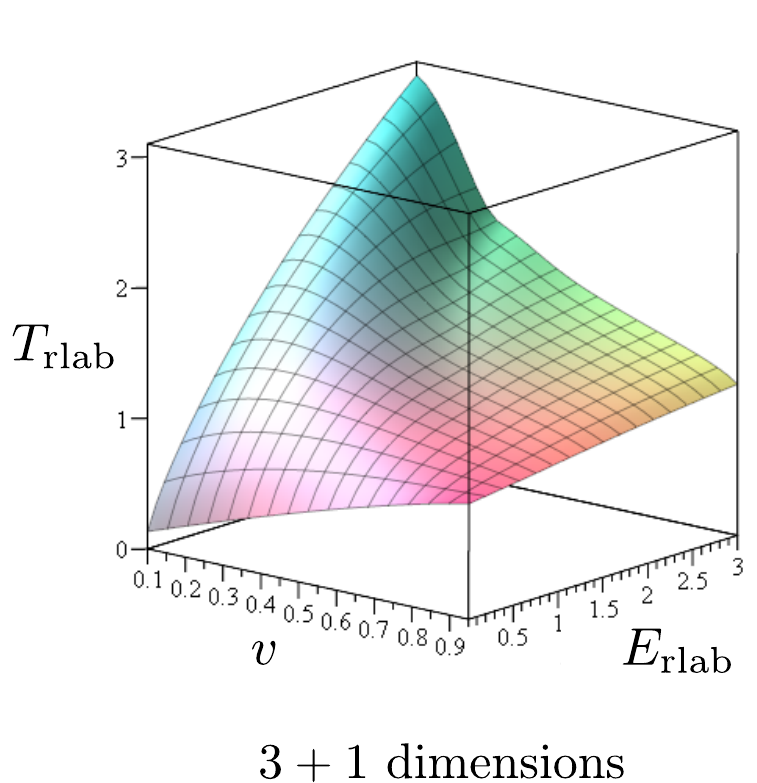} \hspace{2ex} \includegraphics[width=0.48\textwidth]{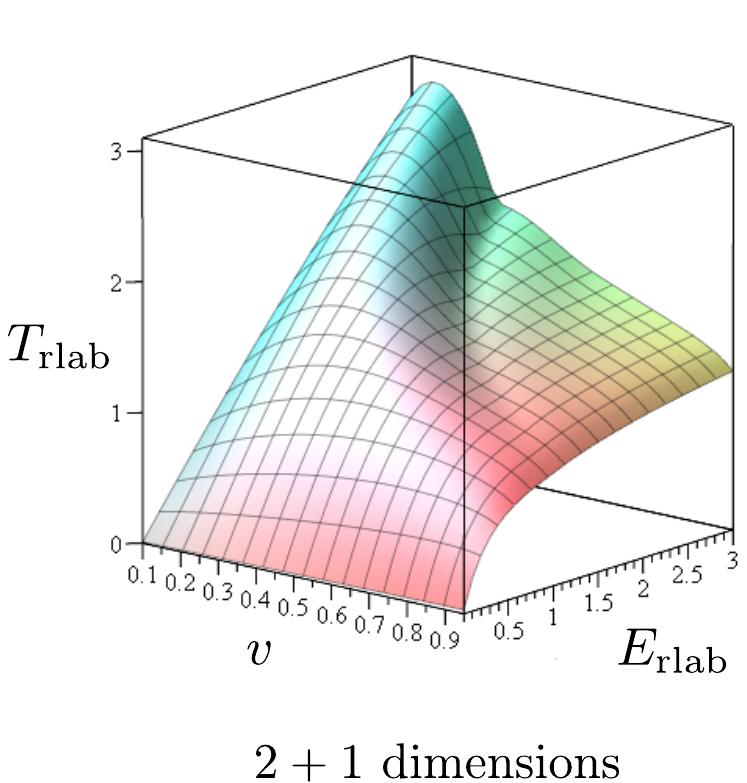}
\caption{Analogue spacetime $T_{\text{rlab}} := {\hat T}_{\text{circ}}/{\hat T}_{\text{lin}}$ 
as a function of $v$ and $E_{\text{rlab}} := {\hat E}/{\hat a}$, 
for $0.1 \le v \le 0.95$ and $0.1 \le E_{\text{rlab}} \le 3$. 
The plotting range was again chosen for numerical stability, 
avoiding small and large values of $v$ and small and large values of $E_{\text{rlab}}$. 
Left in $3+1$ dimensions; right in $2+1$ dimensions. 
The data is as in Figure~\ref{fig:perspectiveTrat}, 
and $T_{\text{rlab}} = T_{\text{rat}}$, 
but $E_{\text{rlab}} = \gamma E_{\text{red}}$.
In the limit $E_{\text{red}}\to0$, outside the plotted range, it is again the case that 
the $3+1$ graph tends to a nonzero value while  
the $2+1$ graph has a significant drop, tending to zero linearly in~$E_{\text{red}}$. 
In the near-sonic limit $v\to1$, outside the plotted range, 
the $3+1$ graph tends to the constant value $\pi/(2\sqrt{3}) \approx 0.9$, 
as seen from~\eqref{eq:Ttratbec-ultrarel-small}, 
but the $2+1$ graph drops to zero proportionally to $-1/\ln(1-v^2)$, 
as seen from~\eqref{eq:Tratbec-2+1:ultrarel-small}; within the plotted range, 
this drop shows as incipient for $0.9 \lesssim v \le 0.95$.\label{fig:perspectiveThatrat}} 
\end{figure}

\begin{figure}[p]
\centering
\includegraphics[width=0.65\textwidth]{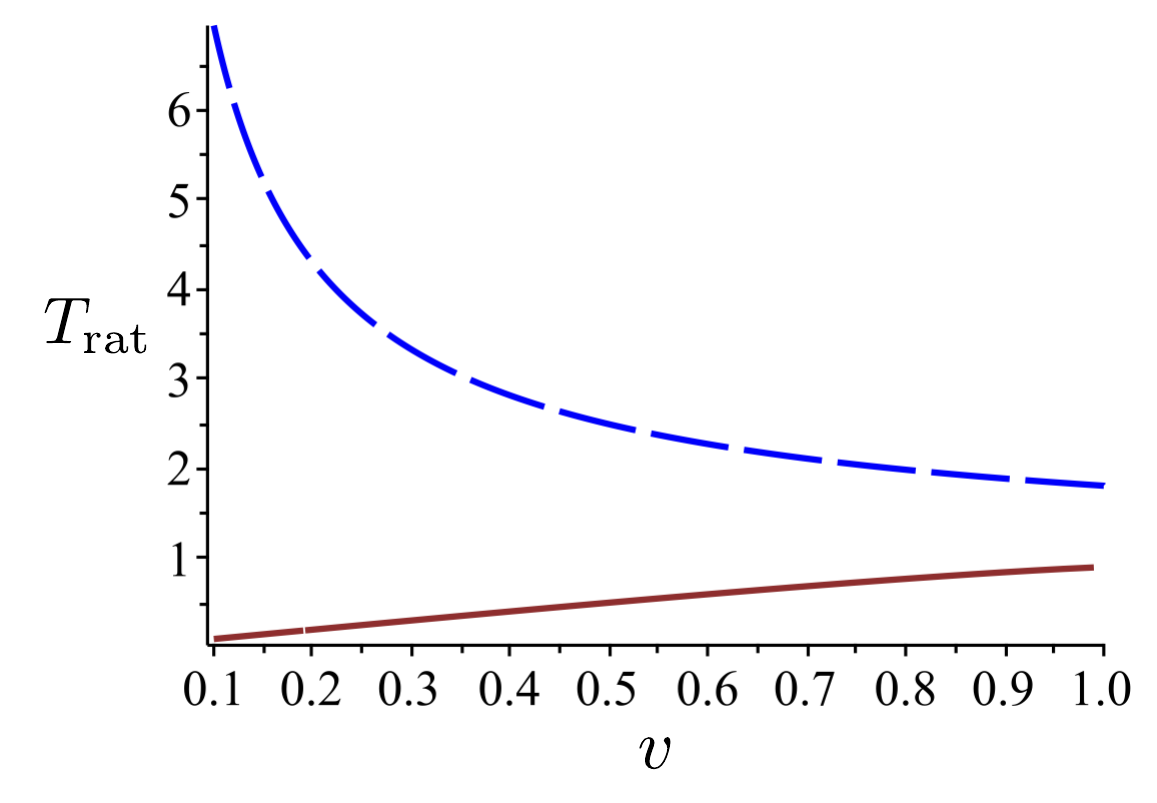}
\caption{Relativistic spacetime 
$T_{\text{rat}} := T_{\text{circ}}/T_{\text{lin}}$ as a function of~$v$ in the limits of large and small~$|E|$, 
showing the continuation of the Figure \ref{fig:perspectiveTrat} plots to these limits. 
The dashed (blue) curve shows the large $|E|$ limit, in both $3+1$ and $2+1$ dimensions, evaluated from \eqref{eq:detailedbalance-temperature} with~\eqref{eq:T/Tlin:largegap}. 
The solid (brown) curve shows the small $|E|$ limit in $3+1$ dimensions, evaluated from~\eqref{eq:T/Tlin:zerogap}. 
In $2+1$ dimensions the small $|E|$ limit vanishes, as seen from the analytic formula~\eqref{eq:2+1-T-smallgap}. 
\label{fig:largeandsmallgap}}
\end{figure}

\begin{figure}[p]
\centering
\includegraphics[width=0.65\textwidth]{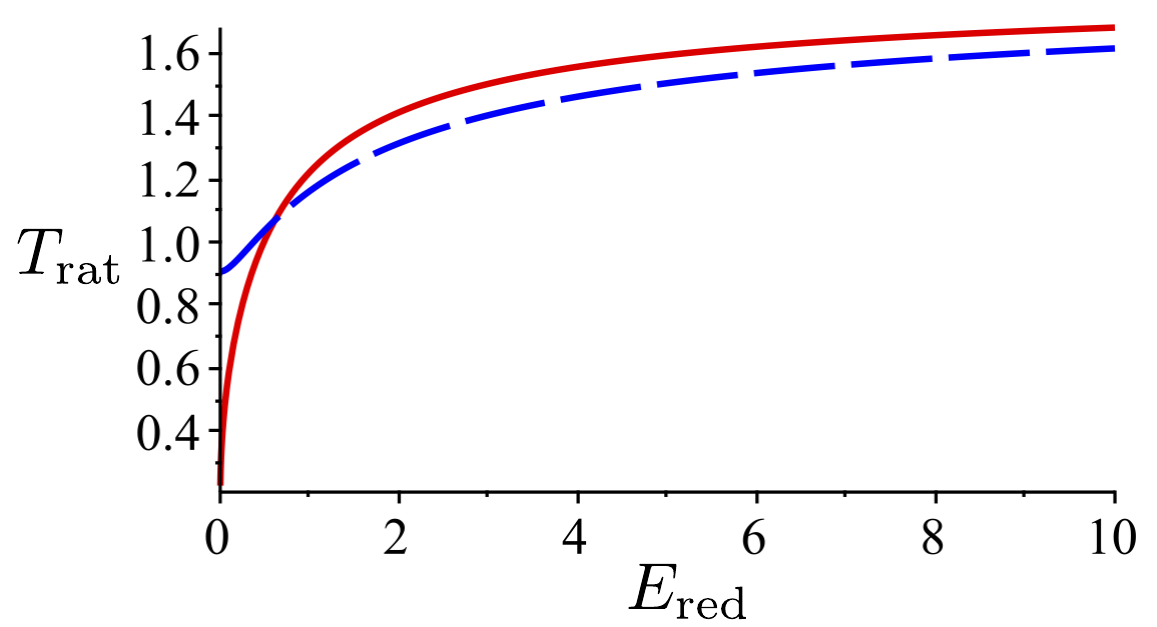}
\caption{Relativistic spacetime 
$T_{\text{rat}} := T_{\text{circ}}/T_{\text{lin}}$ as a function of $E_{\text{red}} := E/a$ in the ultrarelativistic limit, $v\to1$,
showing the continuation of the Figure \ref{fig:perspectiveTrat} plots to this limit. 
The dashed (blue) curve is for $3+1$ dimensions, evaluated from~\eqref{eq:T-ultrarel}, interpolating between $\pi/\sqrt{3} \approx 1.8$ as $E_{\text{red}} \to\infty$ and $\pi/(2\sqrt{3}) \approx 0.9$ as $E_{\text{red}} \to0$, 
as previously found in~\cite{unruhcirclong}. 
The solid (red) curve is for $2+1$ dimensions, evaluated from~\eqref{eq:T-2+1-ultrarel}, 
interpolating between $\pi/\sqrt{3} \approx 1.8$ as $E_{\text{red}} \to\infty$ 
and 
$0$ as $E_{\text{red}} \to0$, showing the falloff proportional to 
$1/\ln(1/E_{\text{red}})$ \eqref{eq:T-2+1:ultrarel-small} as $E_{\text{red}} \to0$. \label{fig:ultrarel}}
\end{figure}

\section{Numerical results\label{sec:numerical}} 

For the relativistic spacetime system, perspective plots of $T_{\text{circ}}/T_{\text{lin}} = 2\pi T_{\text{circ}}/a$ 
as a function of $v$ and $E/a$ are shown 
Figure~\ref{fig:perspectiveTrat}, both in $3+1$ dimensions and in $2+1$ dimensions. 
The plots confirm that in the ultrarelativistic limit 
$T_{\text{circ}}/T_{\text{lin}}$ is close to the linear motion value $1$ for all 
$E/a$ in $3+1$ dimensions and for $|E|/a \gtrsim 0.5$ in $2+1$ dimensions. 
For $|E|/a \ll 1$, the $2+1$ temperature is however significantly lower 
than the $3+1$ temperature for all~$v$. 

For the analogue spacetime system, the corresponding plots of ${\hat T}_{\text{circ}}/{\hat T}_{\text{lin}}$
as a function of $v$ and ${\hat E}/{\hat a}$ are shown in Figure~\ref{fig:perspectiveThatrat}, 
both in $3+1$ dimensions and in $2+1$ dimensions. 
In $2+1$ dimensions there is again a significant drop at small~${\hat E}/{\hat a}$. 

For the relativistic spacetime system, 
the plots in Figure \ref{fig:perspectiveTrat} are complemented by the 
large and small $|E|$ limits shown in Figure~\ref{fig:largeandsmallgap}, 
and by the ultrarelativistic limit at fixed $E/a$ shown in Figure~\ref{fig:ultrarel}.

\section{Conclusions and experimental upshots\label{sec:conclusions}}

Motivated by recent proposals to observe the circular motion Unruh effect in a condensed matter analogue spacetime system \cite{Retzker-circular,unruh-rqin19talk,Gooding:2020scc}, we have presented a detailed comparison of the linear acceleration Unruh temperature $T_{\text{lin}}$ and the circular acceleration Unruh temperature~$T_{\text{circ}}$, for a massless scalar quantum field in its Minkowski vacuum state, in spacetime dimensions 3+1 and 2+1. We considered both a genuine relativistic spacetime system and an analogue spacetime laboratory implementation, the difference being that the laboratory system has no time dilation, so that the systems are mapped to each other by scaling the energies by the time dilation gamma-factor. We probed the field by a pointlike Unruh-DeWitt detector, linearly coupled to the field, working in the limit of weak interaction and long interaction time \cite{Unruh:1976db,DeWitt:1979}, neglecting the detector's back-action on the field. We obtained analytic results in several limits and provided numerical results for the interpolating regions.

An expected outcome was that the highest temperatures, and hence the best experimental prospects, are at the ultrarelativistic limit in the relativistic system and at the 
near-sonic
limit in the analogue spacetime system, 
both in 3+1 dimensions and in 2+1 dimensions. 
In the special case of the 3+1 relativistic system, we in particular confirmed the results obtained previously 
in~\cite{unruhcirclong}. 
An unexpected outcome was, however, that in 2+1 dimensions $T_{\text{circ}}$ contains a 
logarithmic suppression factor in certain limits, 
including the near-sonic 
limit in the analogue spacetime 
system: while the analogue spacetime temperature 
grows without bound in the near-sonic limit in both 3+1 and 2+1 dimensions, 
the growth is slower in 2+1 dimensions. 
This suppression could help prospective analogue spacetime experiments with an effective spacetime dimension 2+1 to stay longer within the regime in which our linear perturbation theory analysis remains valid. 

While we leave it to future work to address effects due to other phenomena that will inevitably be present in experimental implementations, including 
finite size~\cite{Levin-Peleg-Peres,Davies:1996ks,Gutti:2010nv}, 
finite interaction time~\cite{Fewster:2016ewy}, 
nonzero ambient temperature~\cite{Hodgkinson:2014iua}, 
dispersion relation nonlinearity and Lorentz-noninvariance~\cite{Gutti:2010nv,Stargen:2017xii,Louko:2017emx}, 
and the detector's back-action on the field
\cite{Lin:2006jw,Moustos:2016lol,Sokolov:2018nmu}, 
we shall end here with a comment on the classical versus quantum nature of the circular motion Unruh effect. 

The Unruh-DeWitt detector analysed in this paper is a genuinely quantum detector 
coupled to a genuinely quantum field. 
It was observed in \cite{Leonhardt:2017lwm} that some properties of 
similar quantum systems can be modelled by classical Gaussian 
noise when the quantum phenomena 
are analysed in terms of Bogoliubov coefficient techniques. 
In our circular motion setting, 
where Bogoliubov coefficients are not an applicable tool \cite{Levin-Peleg-Peres,Davies:1996ks,Gutti:2010nv}, 
could the response of a localised Unruh-DeWitt detector in circular motion be modelled and simulated by classical Gaussian noise? 

The key observation here is that the two-point correlation function of a Gaussian noise 
is the real part of the quantum field's Wightman function. 
A measurement of the two-point correlation function of classical Gaussian noise 
(say, of thermal fluctuations in a classical fluid) along a circular trajectory 
would hence allow us to recover the part of the response function ${\mathcal F}(E)$ 
that is even in the energy~$E$, 
but not the part that is odd in~$E$. 
Now, the amount of information about ${\mathcal F}$ that is encoded in the even part 
depends on the spacetime dimension. 
In $3+1$ dimensions, 
the even part of ${\mathcal F}$ \eqref{eq:F-combined} contains most of the information of interest, 
and in particular it contains all of the dependence on the parameters of the orbit. 
In $2+1$ dimensions, by contrast, 
the even part of ${\mathcal F}$ \eqref{eq:F2+1-first-split}
is a universal additive constant, independent of the parameters 
of the orbit and even independent of~$E$, while 
all the information of interest is contained in the odd part. 
This implies that a laboratory experiment to observe the fluctuations responsible for the 
$(2+1)$-dimensional circular motion Unruh effect by a localised Unruh-DeWitt type detector 
will need to be a genuinely quantum experiment.

\section*{Acknowledgments}

We thank 
Benito Ju\'arez-Aubry, 
Kinjalk Lochan, 
Ralf Sch\"utzhold 
and 
Pierre Verlot 
for discussions and 
correspondence, and an anonymous referee for helpful presentational suggestions. 
This work originated at a June 2018 Unruh effect workshop 
at the University of Nottingham, supported by 
SW's Foundational Questions Institute (FQXi) Mini-Grant (FQXi-MGB-1742 ``Detecting Unruh Radiation'') and, 
in part, was made possible by 
United Kingdom Research and Innovation (UKRI) Science and Technology Facilities Council (STFC) grant ST/S002227/1 
``Quantum Sensors for Fundamental Physics.'' 
JL and WGU thank the organisers of the 
Relativistic Quantum Information 2019 School and Conference, 
Tainan, Taiwan, 25 May -- 1 June 2019, 
where part of this work was done. 
SE acknowledges support through the Wiener Wissenschafts- und TechnologieFonds (WWTF) project No MA16-066 (``SEQUEX''), and funding from the European Union's Horizon 2020 research and innovation programme under the Marie Sk\l{}odowska-Curie grant agreement No 801110 and the Austrian Federal Ministry of Education, Science and Research (BMBWF) from an 
Erwin Schr\"odinger Center for Quantum Science and Technology
(ESQ) fellowship. 
SE and SW acknowledge support from the Engineering and Physical Sciences Research Council Project Grant (EP/P00637X/1). 
JL and SW acknowledge partial support by 
Science and Technology Facilities Council 
(Theory Consolidated Grant ST/P000703/1). 
JS acknowledges support from the European Research Council Advanced Grant QuantumRelax.
WGU acknowledges support by NSERC Canada (Natural Science and Engineering Research Council), 
the Hagler Fellowship from HIAS (Hagler Institute for Advanced Study) at Texas A\&M University, 
CIfAR (Canadian Institute for Advanced Research), and
the Humboldt Foundation. 
SW acknowledges support provided under the Paper Enhancement Grant at the University of Nottingham, 
the Royal Society University Research Fellow (UF120112), 
the Nottingham Advanced Research Fellow (A2RHS2) and 
the Royal Society Enhancement Grant (RGF/EA/180286).  
This manuscript reflects only the authors' view, 
the European Union Agency is not responsible for any use that 
may be made of the information contained herein.

\appendix 

\section{Appendix: Zeroes of an auxiliary function\label{app:auxiliary-zeroes}}

In this appendix we locate and parametrise the zeroes of the function $f$ defined by 
\begin{subequations}
\begin{align}
f(z) &:= f_+(z) f_-(z)
\ , 
\\
f_{\pm}(z) &:= \frac{z}{v} \pm \sin z
\ , 
\end{align}
\end{subequations}
where $0 < v < 1$ and $z$ is a complex variable. 
Previous discussion is given in~\cite{unruhcirclong,Muller:1995vk}. 

Consider first the real zeroes. 
Each $f_\pm(z)$ has a simple zero at 
$z=0$ and no other real zeroes. 
Hence the only real zero of $f$ is a double zero at $z=0$. 
Note that $f(z)>0$ for all real nonvanishing~$z$. 

To consider the non-real zeroes, we parametrise 
$v$ by 
\begin{align}
v = \frac{\alpha_0}{\sinh\alpha_0}
\ , 
\label{eq:v-param}
\end{align}
where $\alpha_0>0$. As $f$ is even, it suffices to give the zeroes in the upper half-plane. 

We write the zeroes in the upper half-plane as $z_n = i(\alpha_n + i \beta_n)$, 
where $n\in\BbbZ$, $\alpha_n>0$ and $\beta_n\in\BbbR$. 
$\alpha_0$ is given by \eqref{eq:v-param} and $\beta_0=0$. 
For $n\ne0$, $\alpha_n$ is the unique positive zero of the function 
\begin{align}
g_n(\alpha) = 
- \sqrt{1 - \left( \frac{\sinh\alpha_0}{\alpha_0}\frac{\alpha}{\sinh\alpha}\right)^2} \frac{\alpha_0}{\sinh\alpha_0} \cosh\alpha
+ \arccos\!\left( \frac{\sinh\alpha_0}{\alpha_0}\frac{\alpha}{\sinh\alpha}\right)
+ |n| \pi 
\label{eq:g-def}
\end{align}
and 
\begin{align}
\beta_n = \sgn(n) \! \left[ \arccos\!\left( \frac{\sinh\alpha_0}{\alpha_0}\frac{\alpha_n}{\sinh\alpha_n}\right)
+ |n| \pi\right]
\ . 
\label{eq:betan-def}
\end{align}
Even $n$ give the zeroes of $f_-$ and odd $n$ give the zeroes of $f_+$. 
All these zeroes are simple. 

The zeroes satisfy 
\begin{subequations}
\begin{align}
&0 < \alpha_0 < \alpha_{\pm1} < \alpha_{\pm2} < \cdots
\ ,  
\label{eq:alphasequence}
\\
&0 = \beta_0 < |\beta_{\pm1}| < |\beta_{\pm2}| < \cdots
\ . 
\end{align}
\end{subequations}
At $|n|\to\infty$ with fixed~$v$, the leading asymptotics is 
\begin{subequations}
\label{eq:alphabeta-n-asymptotics}
\begin{align}
\alpha_n & \sim \ln \! \left( \frac{(2|n| + 1) \pi \sinh\alpha_0}{\alpha_0} \right) 
\ ,  
\\
\beta_n &\sim \sgn(n) \bigl(|n| + {\textstyle{\frac12}} \bigr) \pi
\ . 
\end{align}
\end{subequations}

In the limit $v\to1$, 
\eqref{eq:v-param} gives $\alpha_0\to0$ 
and
\begin{subequations}
\label{eq:v-versus-alpha0:as}
\begin{align}
v &= 1 - \tfrac16 \alpha_0^2 + O(\alpha_0^4)
\ , 
\\
\alpha_0^2 &= 6 (1- v) + O\bigl((1-v)^2\bigr)
\ . 
\end{align}
\end{subequations}
For $n\ne0$, $\alpha_n$ and $\beta_n$ tend in this limit to the nonzero values obtained from 
\eqref{eq:g-def}
and
\eqref{eq:betan-def}
after the replacement $\frac{\sinh\alpha_0}{\alpha_0} \to 1$.

\section{Appendix: 2+1 small gap limit\label{app:2+1-smallgap}}

In this appendix we verify the 2+1 small gap property \eqref{eq:Fcorr2+1-smallgap} stated 
in subsection~\ref{sec:2+1-smallgap}. 

Starting from \eqref{eq:F2+1-corr}, 
we write ${\mathcal F}^{\text{corr}}(E) = {(2\pi\gamma)}^{-1} P\bigl(\frac{2ER}{\gamma v}\bigr)$, 
where 
\begin{align}
P(b) &= \int_0^\infty dz \sin(bz) \left( \frac{\gamma}{z} - \frac{1}{\sqrt{z^2 - v^2 \sin^2\!z}} \right) 
\notag
\\
& = \int_0^\infty dz \sin(bz) \left( \frac{\gamma-1}{z} + \frac{1}{z} - \frac{1}{\sqrt{z^2 - v^2 \sin^2\!z}} \right) 
\notag
\\
& = \frac{\pi (\gamma-1)}{2} \sgn(b)
+ b \int_0^\infty dz \, \frac{\sin(bz)}{bz} \left( 1 - \frac{z}{\sqrt{z^2 - v^2 \sin^2\!z}} \right) 
\notag
\\
& = \frac{\pi (\gamma-1)}{2} \sgn(b) 
+ b \int_0^\infty dz \left( 1 - \frac{z}{\sqrt{z^2 - v^2 \sin^2\!z}} \right) 
\ + o(b)
\ . 
\label{eq:Fcorr2+1-smallgap-Pform}
\end{align}
In \eqref{eq:Fcorr2+1-smallgap-Pform}
we have added and subtracted a multiple of $\sin(bz)/z$ under the integral, 
used the standard integral $\int_0^\infty dz \, \frac{\sin(bz)}{z} = \tfrac12 \pi \sgn(b)$, 
and in the last step used a dominated convergence argument to take the limit under the integral. 
This establishes~\eqref{eq:Fcorr2+1-smallgap}.

\section{Appendix: 2+1 ultrarelativistic limit with fixed $E$\label{app:2+1-ultrarel-fixedE}}

In this appendix we verify the $2+1$ fixed $E$ ultrarelativistic 
limit property \eqref{eq:Fodd-2+1-fixedE-as} stated in subsection~\ref{sec:2+1-ultrarel-fixedE}. 

From \eqref{eq:F2+1-odd-def} we have 
\begin{align}
{\mathcal F}^{\text{odd}}(E)
= - \frac{1}{2\pi} \sigma_\gamma (2ER)
\ , 
\label{eq:F-2+1-odd-intermsof-sigma}
\end{align}
where 
\begin{align}
\sigma_\gamma (b) = 
\int_0^\infty dx \, \frac{\sin(bx)}{\sqrt{\gamma^2 x^2  - \sin^2 \Bigl(\sqrt{\gamma^2-1} \, x\Bigr)}}
\ , 
\label{eq:sigma-gamma-def}
\end{align}
after the change of variables $z = \gamma v x$. 
We shall show that 
\begin{align}
\sigma_\gamma(b) = \frac{\pi}{2\gamma} \sgn(b) + o(1/\gamma)
\end{align}
when $\gamma\to\infty$ with fixed~$b$. 
\eqref{eq:Fodd-2+1-fixedE-as} then follows from \eqref{eq:F-2+1-odd-intermsof-sigma}. 

Let $b\ne0$ be fixed. Using the standard integral $\int_0^\infty dx \, \frac{\sin(bx)}{x} = \frac{1}{2} \pi \sgn(b)$, 
we rearrange 
\eqref{eq:sigma-gamma-def} as 
\begin{align}
\sigma_\gamma(b) = \frac{\pi}{2\gamma} \sgn(b)
+ \frac{1}{\gamma} I_\gamma(b)
\ , 
\end{align}
where 
\begin{align}
I_\gamma(b) = \int_0^\infty dx \, \frac{\sin(bx)}{x}
\left\{
\left[1 - \frac{\sin^2 \Bigl(\sqrt{\gamma^2-1} \, x\Bigr)}{\gamma^2 x^2 }\right]^{\!-1/2}
- 1 
\right\}
\ . 
\label{eq:Igamma-def}
\end{align}
We need to show that $I_\gamma(b)\to0$ as $\gamma\to\infty$. 

Let $M>1$ be a constant, and let $\gamma$ be so large that $\pi/ \sqrt{\gamma^2-1} < M$.
Let $I^{(1)}_\gamma$, $I^{(2)}_\gamma$ and $I^{(3)}_\gamma$ 
denote respectively the contributions to \eqref{eq:Igamma-def} from 
$0< x < \pi/ \sqrt{\gamma^2-1}$, 
$\pi/ \sqrt{\gamma^2-1}< x < M$ 
and 
$M < x < \infty$. We consider each in turn. 

In $I^{(3)}_\gamma$, the integrand goes pointwise to zero as $\gamma\to\infty$ and is bounded in absolute value by the integrable function $A_3/x^3$ where $A_3$ is a $\gamma$-independent constant. Hence $I^{(3)}_\gamma \to0$ as $\gamma\to\infty$ by dominated convergence. 

In $I^{(2)}_\gamma$, we first write the integral to be over the $\gamma$-independent interval $0< x < M$ by defining the integrand to have the value zero for $0 < x \le \pi/ \sqrt{\gamma^2-1}$. An elementary argument then shows that the integrand goes pointwise to zero as $\gamma\to\infty$ and is bounded in absolute value by a $\gamma$-independent constant. 
Hence $I^{(2)}_\gamma \to0$ as $\gamma\to\infty$ by dominated convergence. 

In $I^{(1)}_\gamma$, changing the integration variable by $x = z/\sqrt{\gamma^2-1}$ gives 
\begin{align}
I^{(1)}_\gamma(b) = \int_0^\pi dz \, 
\frac{\sin\Bigl(\frac{bz}{\sqrt{\gamma^2-1}}\Bigr)}{\Bigl(\frac{z}{\sqrt{\gamma^2-1}}\Bigr)}
\left\{ 
\frac{1}{\sqrt{\gamma^2 - 1}} 
\left[1 - \left(\frac{\gamma^2 - 1}{\gamma^2}\right)\frac{\sin^2z}{z^2}\right]^{\!-1/2}
- 
\frac{1}{\sqrt{\gamma^2 - 1}}
\right\} 
\ . 
\label{eq:I3-mod}
\end{align}
In \eqref{eq:I3-mod}, the integrand goes to zero pointwise at each positive $z$ as $\gamma\to\infty$, and an elementary argument using the properties of $\frac{\sin z}{z}$ shows that the integrand is bounded in absolute value by a $\gamma$-independent constant. 
Hence $I^{(1)}_\gamma \to0$ as $\gamma\to\infty$ by dominated convergence. 

This completes the argument.

\section{Appendix: 2+1 ultrarelativistic limit with fixed $E/a$\label{app:2+1-ultrarel-E/a-fixed}}

In this appendix we verify the 2+1 ultrarelativistic limit properties stated 
in subsection~\ref{sec:2+1-ultrarel}. 

\subsection{Taking the limit}

There are two ways to obtain the limit. 

One way is to start from \eqref{eq:F2+1-odd-def} and write $z = y/\gamma$, giving 
\begin{align}
{\mathcal F}^{\text{odd}}(E)
= 
- \frac{1}{2\pi} 
\int_0^\infty dy \, 
\frac{\sin\!\left(\frac{2ER}{\gamma^2 v} y\right)}{\sqrt{\gamma^2 y^2 - \gamma^2(\gamma^2-1)\sin^2 \! (y/\gamma)}} 
\ .  
\label{eq:F2+1odd-y}
\end{align}
Now take $\gamma\to\infty$ with $2ER/(\gamma^2 v)$ fixed. 
The function under the square root in \eqref{eq:F2+1odd-y} has the pointwise limit $y^2 \bigl(1+\frac13 y^2 \bigr)$, 
and taking the limit under the integral can be justified by breaking the domain into half-periods of the sine, 
combining pairwise the contributions from adjacent intervals, 
and invoking a dominated convergence argument to take the limit under the sum. 
(Evidence for the existence of a dominating summable function was obtained numerically from Maple.) 
Writing finally $y = \sqrt3 \,x$, we obtain~\eqref{eq:Fodd-2+1-limit}. 

Another way is to start from \eqref{eq:F0corr:2+1-combined} 
and take $\alpha_0\to0$ with $2ER\alpha_0/(\gamma v)$ fixed. 
The contribution from ${\mathcal F}^{\text{corr}}_{+}$ 
vanishes by a dominated convergence argument in~\eqref{eq:F+corr:2+1}. 
For ${\mathcal F}^{\text{corr}}_{0}$, writing $\alpha = \alpha_0 y$ in 
\eqref{eq:F0corr:2+1} gives 
\begin{align}
{\mathcal F}^{\text{corr}}_0(E) &= 
\frac{\sgn(E)}{2\pi \alpha_0\gamma v} \int_{1}^\infty 
dy \, 
\frac{\exp\!\left(-\frac{2|E|R \alpha_0}{\gamma v} y\right)}{\sqrt{\alpha_0^{-4}\left(\sinh^2 \! (\alpha_0 y) - y^2 \sinh^2 \!\alpha_0 \right)}} 
\ . 
\label{eq:F02+1-y}
\end{align}
The function under the square root in \eqref{eq:F02+1-y} has the pointwise limit $\frac13 y^2 \bigl(y^2 -1\bigr)$ 
and is bounded below by this limit. 
Taking the limit $\alpha_0 \to0$ under the integral is 
hence justified by dominated convergence, with the outcome~\eqref{eq:Fcorr-2+1-limit}. 

To verify that the functions
appearing in these limits satisfy~\eqref{eq:H-G-relation}, 
we start from 
\eqref{eq:Gfunc-def} and proceed as in Appendix C of~\cite{Hodgkinson:2014iua}, 
\begin{align}
\sgn(q) G(q) &= \frac12 \int_{-\infty}^\infty dx \, \frac{\sin(|q|x)}{x \sqrt{1+x^2}}
\notag
\\
&= \frac{\pi}{2} 
+ \frac12 \int_{-\infty}^\infty dx \sin(|q|x) \left(\frac{1}{x \sqrt{1+x^2}} - \frac{1}{x} \right) 
\notag
\\
&= \frac{\pi}{2} 
- \frac{i}{2} \int_{-\infty}^\infty dx \, e^{i|q|x} \left(\frac{1}{x \sqrt{1+x^2}} - \frac{1}{x} \right) 
\notag
\\
&= \frac{\pi}{2} - \frac{i}{2} \int_C dz \, \frac{e^{i|q|z}}{z \sqrt{1+z^2}}
\notag
\\
&= \frac{\pi}{2} -  \int_1^\infty dy \, \frac{e^{-|q| y}}{y \sqrt{y^2-1}}
\notag
\\
&= \frac{\pi}{2} -  \sgn(q) H(q)
\ , 
\end{align}
where the contour $C$ is along the real axis from $-\infty$ to $\infty$ 
except for passing the pole at $z=0$ in the upper half-plane. 
$C$~is then deformed to the upper half-plane, encircling the branch point at $z=i$ and running on
both sides of the cut at $z=iy$ with $y>1$. 
The last equality uses~\eqref{eq:Hfunc-def}. 
This gives~\eqref{eq:H-G-relation}. 

\subsection{Small argument form of $G$ \eqref{eq:Gfunc-def}}

To find the small argument form of $G$~\eqref{eq:Gfunc-def}, 
we introduce a positive constant $M$ and split 
\eqref{eq:Gfunc-def} as 
\begin{subequations}
\begin{align}
G(q) &= q \bigl( I_>(q) + I_<(q) \bigr)
\ , 
\\
I_>(q) &= \int_M^\infty dz \, \frac{\sin z}{z^2\sqrt{1 + (q/z)^2}}
\ , 
\label{eq:I-large-def}
\\
I_<(q) &= \int_0^M dz \, \frac{\sin z}{z\sqrt{z^2 + q^2}}
\ , 
\label{eq:I-small-def}
\end{align}
\end{subequations}
recalling that $q\ne0$ by assumption and 
using the substitution $x = z/q$. 

From \eqref{eq:I-large-def} we have 
\begin{align}
I_>(q) &= \int_M^\infty dz \, \frac{\sin z}{z^2}
\ \ + O(q^2)
\notag
\\
&= \frac{\sin M}{M} + \int_M^\infty dz \, \frac{\cos z}{z}
\ \ + O(q^2)
\ , 
\label{eq:I-large-exp}
\end{align}
first expanding in $q$ and then integrating by parts. 
From \eqref{eq:I-small-def} we have 
\begin{align}
I_<(q) 
&= \int_0^M \frac{dz}{\sqrt{z^2 + q^2}} 
+ 
\int_0^M \frac{dz}{\sqrt{z^2+q^2}} \left( \frac{\sin z}{z} -1 \right) 
\notag
\\
&= \arsinh(M/|q|)
+ 
\int_0^M \frac{dz}{z} \left( \frac{\sin z}{z} -1 \right) 
+ o(1) 
\notag
\\
&= 
\ln(2M/|q|) + 
\int_0^M \frac{dz}{z} \left( \frac{\sin z}{z} -1 \right) 
+ o(1) 
\ , 
\label{eq:I-small-exp}
\end{align}
first splitting the integrand, then evaluating the elementary integral of 
the first term and taking the limit in the second term by dominated convergence, 
and finally expanding the arsinh. 

Combining \eqref{eq:I-large-exp} and \eqref{eq:I-small-exp} gives
\begin{align}
I_>(q) + I_<(q) 
&= 
\ln(2/|q|) 
+ \int_0^M \frac{dz}{z} \left( \frac{\sin z}{z} -1 \right) 
+ \frac{\sin M}{M} 
+ 
\ln M 
+ \int_M^\infty dz \, \frac{\cos z}{z} 
+ o(1) 
\notag
\\
&= 
\ln(2e^{1-\gamma_E}/|q|) 
+ o(1)
\ , 
\label{eq:I-large+small:exp}
\end{align}
where $\gamma_E$ is Euler's constant, 
and the $M$-independent sum of the individually 
$M$-dependent terms has been evaluated by taking the limit 
$M\to0$ and using the small argument expansion 
of the cosine integral from 6.2.13 in~\cite{dlmf}. 
Hence 
\begin{align}
G(q)
= 
q \ln(2e^{1-\gamma_E}/|q|) 
+ o(q)
\ , 
\end{align}
which is \eqref{eq:G-asymptotics}.

\subsection{Large argument form of $H$ \eqref{eq:Hfunc-def}}

To find the large argument form of $H$~\eqref{eq:Hfunc-def}, 
we first substitute $y = 1 + r^2$ 
and then use the stationary point expansion~\cite{wong}, obtaining 
\begin{align}
H(q) &= \sgn(q) \int_0^\infty dr \, \frac{2 e^{-|q|r^2}}{(1+r^2)\sqrt{2 + r^2}}
\notag
\\
&= 
\sgn(q) 
\sqrt{\frac{\pi}{2|q|}} \, e^{-|q|} \Bigl(1 + O\bigl(|q|^{-1}\bigr) \Bigr)
\ , 
\end{align}
which is~\eqref{eq:H-asymptotics}.

\section{Appendix: Analogue spacetime asymptotics\label{app:analog-spacetime-asymptotics}}

In this appendix we verify the asymptotic 
temperature formulas \eqref{eq:Tbecboth-ultrarel-small}
for the analogue spacetime implementation. 

To verify~\eqref{eq:Tbec-ultrarel-small}, we use \eqref{eq:T-ultrarel-small}, 
which is allowed because the 
$v\to1$ limit \eqref{eq:T-ultrarel} is uniform in~$E/a$. 

To verify~\eqref{eq:Tbec-2+1:ultrarel-small}, we note from \eqref{eq:F2+1-odd-def} 
that 
\begin{align}
{\mathcal F}^{\text{odd}}(E) = - \frac{1}{2\pi} \, \rho_\gamma\bigl(2{\hat E}R/v\bigr)
\ , 
\end{align}
 where 
\begin{align}
\rho_\gamma(b) = \frac{1}{\gamma} \int_0^\infty dz \, \frac{\sin(bz)}{\sqrt{z^2 - (1-\gamma^{-2})\sin^2 \! z}} 
\ . 
\end{align}
Writing $x = \gamma z/\sqrt{3}$ in~\eqref{eq:Gfunc-def}, we hence have 
\begin{align}
\gamma \left( \rho_\gamma(b) - G\bigl(\sqrt{3}\,b/\gamma \bigr) \right) = \int_0^\infty dz \, \sin(bz) 
\left( \frac{1}{\sqrt{z^2 - (1-\gamma^{-2})\sin^2 \! z}} - \frac{1}{z \sqrt{\gamma^{-2}+z^2/3}} \right) 
\ . 
\label{eq:rho-G-diff}
\end{align}
When $\gamma\to\infty$ with fixed $b$, the right-hand side of \eqref{eq:rho-G-diff} tends to~$h(b)$, where 
\begin{align}
h(b) :=  \int_0^\infty dz \, \frac{\sin(bz)}{z} 
\left( \frac{1}{\sqrt{1 - (\sin^2 \!  z)/z^2}} - \frac{\sqrt{3}}{z} \right) 
\ . 
\end{align}
Taking the limit under the integral can be justified by breaking the integral 
to $0<z<1$ and $z>1$, using dominated convergence for $0<z<1$, 
and using arguments similar to those in Appendix \ref{app:2+1-ultrarel-E/a-fixed} for $z>1$. 
Using \eqref{eq:G-asymptotics}, we hence have 
\begin{align}
\gamma \rho_\gamma(b) = 
\sqrt{3} \, b \ln\!\left(\frac{2e^{1-\gamma_E}\gamma}{\sqrt{3} \, |b|} \right) 
+ h(b) + o(1) 
\ , 
\end{align}
so that 
\begin{align}
{\mathcal F}^{\text{odd}}(E) \approx - \frac{\sqrt{3}}{\pi} \, \frac{\ln\gamma}{\gamma} \frac{{\hat E}}{{\hat a}}
\ , 
\label{F2+1:odd:BEC-as}
\end{align}
writing $R\approx 1/{\hat a}$ as $v\to1$. 
\eqref{eq:Tbec-2+1:ultrarel-small} now follows from 
\eqref{F2+1:odd:BEC-as} and~\eqref{eq:F2+1-eo-split}.

\end{document}